\documentclass{ieeeaccess}
\usepackage[utf8]{inputenc}
\usepackage{cite}
\usepackage{float}
\usepackage{amssymb,amsfonts}
\usepackage[utf8]{inputenc}
\usepackage{amsmath}
\usepackage{algorithmic}
\usepackage{textcomp}
\usepackage{subcaption}
\usepackage{caption}
\usepackage{tabularx}
\usepackage{longtable}
\usepackage{graphicx,rotating,booktabs}
\usepackage[verbose]{placeins}
\usepackage{framed,multirow}
\usepackage{cuted}

\begin{document}
\history{Date of publication xxxx 00, 0000, date of current version xxxx 00, 0000.}
\doi{10.1109/ACCESS.2024.0429000}

\title{ULTRA: Urdu Language Transformer-based Recommendation Architecture }
\author{\uppercase{Alishbah Bashir}\authorrefmark{1},
\uppercase{Fatima Qaiser}\authorrefmark{1},
\uppercase{Dr. Ijaz Hussain}\authorrefmark{1}}

\address[1]{Department of Computer and Information Sciences, PIEAS, Nilore, Islamabad 45650, Pakistan.}

\corresp{Corresponding author: Ijaz Hussain (ijazhussain@pieas.edu.pk).}

\begin{abstract}

Urdu being a low resource language lacks effective semantic content recommendation systems especially in the area of personalized news retrieval. The current methods are mostly based on lexical matching or language-agnostic methods, which are not very effective at capturing semantic intent and do not work well when the query length and information requirements vary. This weakness leads to poor relevance and flexibility in Urdu content recommendation. To overcome these obstacles, we propose ULTRA (Urdu Language Transformer-based Recommendation Architecture) which is an adaptive semantic recommendation architecture. ULTRA presents a dual-embedding architecture, which has a query-length sensitive routing scheme that dynamically differentiates between short, intent-oriented queries and longer, context-rich queries. According to a threshold-based decision process, user queries are directed to either title/headline-level or full-content/document level-specific semantic pipelines, which guarantee the right level of semantic granularity during retrieval. The suggested system exploits the transformer-based embeddings and optimized pooling techniques to go beyond the surface-level matching of keywords and allow similarity search based on context. The large-scale experiments performed on a large Urdu news corpus indicate that the suggested architecture is always effective in enhancing the relevance of recommendations in relation to different query types. The findings demonstrate precision of over 90\% as compared to single-pipeline baselines. This underscores the usefulness of query-adaptive semantic alignment to low-resource languages. The results make ULTRA a strong and scalable content recommendation structure, providing useful design knowledge to semantic retrieval systems on low-resource language environments.

\end{abstract}

\begin{keywords}{
Content Recommenders, Urdu Content Recommendation System, Intelligent Information Retrieval, Semantic Search, Urdu NLP, Dual-Embedding Framework}
\end{keywords}

\titlepgskip=-15pt

\maketitle

\section{Introduction}
\label{sec:introduction}

\begin{figure*}[h]
\begin{center}
 \includegraphics[width=0.9\textwidth, trim=90 920 90 120, clip]{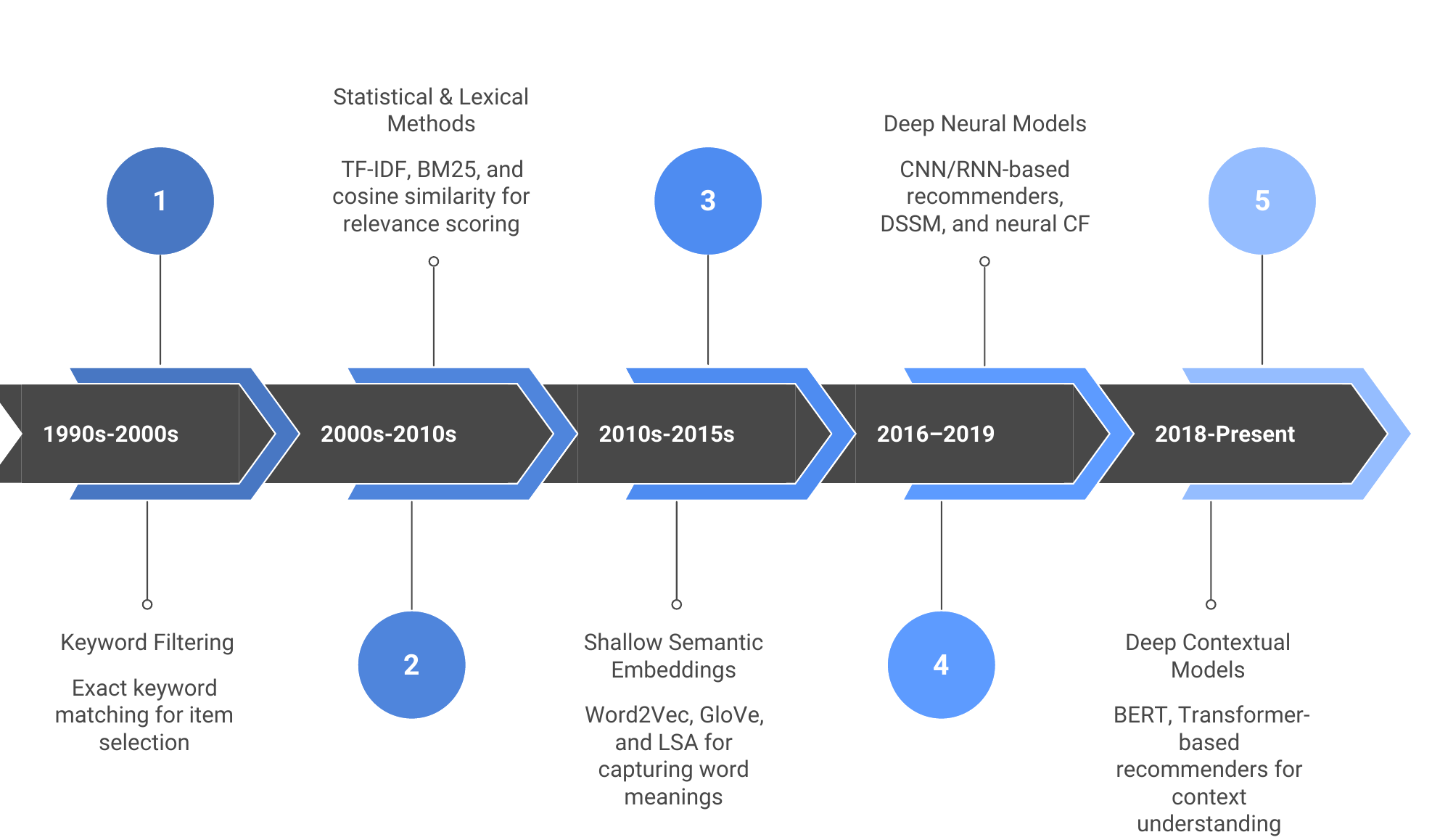}
 \caption{Evolution of text-based recommendation \& retrieval techniques}
  \label{fig:evolution}
  \end{center}
\end{figure*}

Text-based content recommendation systems have become indispensable in the digital era, transforming how users discover and consume information amidst an overwhelming volume of content \cite{1, 2}. By leveraging user behavior, interests, and search queries, these systems curate personalized content feeds, enhancing engagement and information accessibility. At their core, modern systems rely on sophisticated semantic understanding to go beyond simple keyword matching, aiming to grasp user intent and the contextual meaning of content for accurate alignment \cite{3, 4}.  The evolution of text-based recommendation and retrieval techniques from traditional lexical methods to modern semantic approaches is illustrated in Figure \ref{fig:evolution}.

This challenge of semantic recommendation is not uniform; it is heavily modulated by linguistic and computational resource disparities \cite{5}. For high-resource languages like English, advanced transformer-based architectures have achieved remarkable success \cite{6,7,8}. However, for linguistically rich but computationally low-resource languages such as Urdu, the landscape is clearly different. Urdu, with large number of speakers worldwide and written in the complex Nastaliq script, presents unique morphological and syntactic challenges \cite{9,10}. Its context sensitivity, diglossia, and significant code-mixing, particularly with English and Roman Urdu in digital spaces, complicate automated semantic analysis \cite{11,12,13}. This creates a critical barrier to accessing relevant digital content for a vast population.

Existing research aimed at Urdu text-based content recommendation has primarily explored foundational techniques. Early systems and some contemporary approaches rely on traditional lexical methods, such as TF-IDF with cosine similarity, which are fundamentally limited by their dependence on surface-level keyword overlap \cite{14,15}. Acknowledging this limitation, subsequent work has integrated pre-trained transformer models like BERT to generate semantic embeddings, marking a significant step forward \cite{16}. Further innovations have focused on enriching content representation through entity-based techniques, such as wikification and knowledge graphs, aiming to connect content items via named entities and their relationships \cite{17,18}. While these methods improve upon purely lexical systems, they exhibit notable shortcomings that hinder optimal performance in real-world scenarios \cite{24,25,26}.

\begin{figure}[!h]
\begin{centering}
 \includegraphics[width=0.85\linewidth, trim=70 670 110 75, clip]{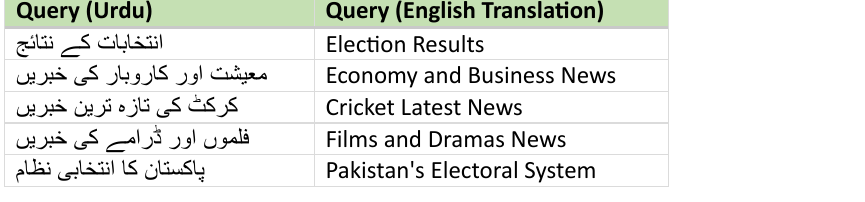}
 \caption{Example of a short, intent-focused Urdu queries}
  \label{fig:Q1}
  \end{centering}
\end{figure}

\begin{figure}[!h]
\begin{centering}
 \includegraphics[width=1\linewidth, trim=10 550 10 70, clip]{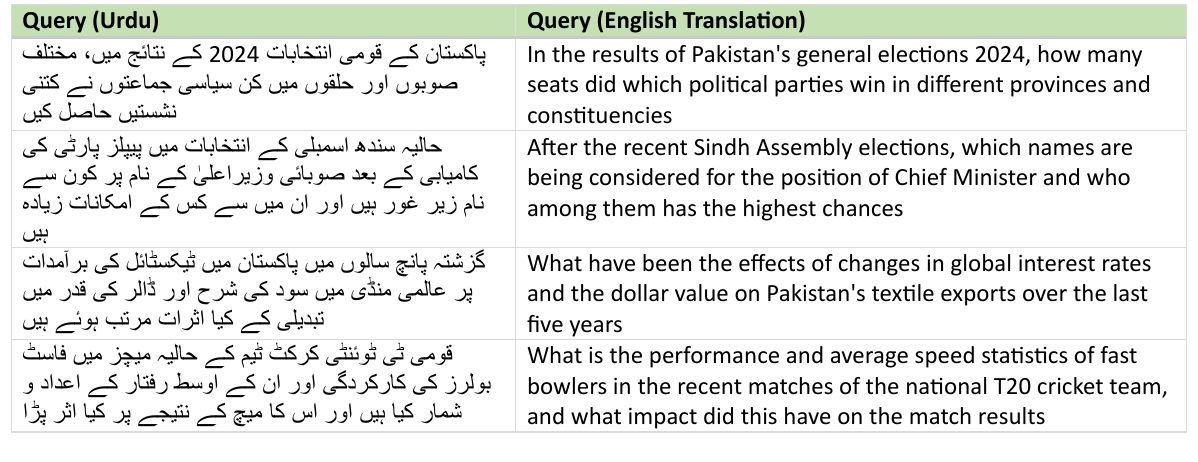}
 \caption{Example of a long, context-rich Urdu queries}
  \label{fig:Q2}
  \end{centering}
\end{figure}

A principal drawback of these existing approaches is their rigid, one-size-fits-all architecture. They typically process all user queries through a single, uniform matching pipeline, whether against headlines, full articles, or entity graphs \cite{19}. This design fails to account for the inherent diversity in user query patterns, which range from short, intent-focused searches as in Figure \ref{fig:Q1}, 

to long, descriptive inquiries seeking specific contextual details as shown in Figure \ref{fig:Q2}. \cite{20}. 

However, remember long queries are not limited to user-entered inputs alone; a detailed definition of long queries, as established in our work, is provided in Section \ref{sec_novelty}.

Consequently, a concise query may be inadequately matched against verbose article bodies, losing precision, while a detailed query may not fully leverage the rich semantics embedded in full-text content, compromising recall \cite{21}. Furthermore, scalability remains a pervasive concern. The computational cost of performing real-time similarity searches on high-dimensional transformer embeddings across large and growing content corpora limits practical deployment \cite{22,23}. There is a clear need for an adaptive system that dynamically tailors its retrieval strategy to the nature of the user's query \cite{27,28,29}.

In order to overcome these severe constraints, we introduce ULTRA (Urdu Language Transformer Based Recommendation Architecture), a new dual-pathway architecture that is used to recommend semantic text-based content in Urdu with a high level of accuracy and adaptability. The main innovation of ULTRA is a smart routing system that dynamically routes user queries to specialized semantic pipelines depending on query properties. This makes sure that the granularity of the informational need of the user is matched with the most suitable level of document representation. With the help of state-of-the-art transformer models optimized to Urdu in an optimized vector search infrastructure, ULTRA can achieve true semantic understanding beyond lexical coincidence, and is efficient enough for large scale use \cite{29}.

\section{Key Contributions}
\label{sec_novelty}

An abstract representation of the ULTRA architecture, depicting the structural organization and functional relationships between its core components is presented in Figure \ref{fig:Art}. Building upon this architectural foundation, the main contributions of this work can be summarized as follows:
 
 \subsection{Use Cases in Text Based Content Recommendation}
 \label{sub_sec_use_case}

When recommending text-based content on content recommendation systems such as news websites, e-commerce sites, blog platforms, and digital libraries, two primary use cases emerge: (1) users enter a query into a search bar, triggering recommendations aligned with the query; and (2) users are engaged with a specific webpage, such as reading a news article, blog post, or product description on an e-commerce site, where related content is suggested below the current page.

Our proposed architecture accommodates both scenarios. In the case of a user entering a query via the search bar that typically consists of keywords and thus classified as a short query, the input is routed to a pipeline optimized for short queries. Conversely, when recommending content, based on the webpage the user is currently viewing, the text on that specific web page serves as the query, which falls under the long query category. Recommendations are then generated relative to the topic the user is actively engaged with. However users can also type long queries in search bar; such inputs would similarly be directed to the long-query-optimized pipeline. To distinguish between short and long queries, we establish a predefined threshold value. The methodology for selecting and determining this threshold is detailed in section Section \ref{sub_sub_sec_adaptive_routing}.

\subsection{Proposed Dual-Pipeline Solution for Identified Use Cases}
 \label{sub_sec_dual_pipe}

The use cases described in Section \ref{sub_sec_use_case} motivate the need for different processing strategies for short and long queries. Our experimental analysis in Section \ref{sec_res-discussion} revealed that short queries, composed of a few keywords, exhibit poor precision when matched against embeddings of lengthy content (e.g., complete articles or blog posts). However, matching these queries against headline embeddings, which are comparable in length and linguistic structure, yielded substantially higher precision.

These observations drove our dual-pipeline architectural design. Rather than matching compact queries directly against full document embeddings, we propose an alternative strategy: (1) encode content semantics via headline embeddings, (2) retrieve relevant headlines based on query-headline similarity, and (3) recommend the complete documents associated with top-ranked headlines. This approach exploits the semantic and structural correspondence between query brevity and headline conciseness. Hence, it improves recommendation precision for Urdu text corpora.

\subsection{Summary of Main Contributions}
\begin{itemize}
    \item We developed ULTRA, a novel adaptive recommendation architecture that introduces a query-length threshold-based routing mechanism, as established in  Section \ref{sub_sec_use_case} and \ref{sub_sec_dual_pipe}, to dynamically optimize semantic matching, a novel approach for Urdu text-based content recommendation.
    \item We conduct empirical evaluation of embedding generation and pooling strategies, establishing optimal configurations for representing both title-level and full document-level Urdu content.
    \item We implement and rigorously compare multiple dimensionality reduction techniques as well as reduced dimension sizes. This helps us identify the most effective method and dimension size for compressing semantic embeddings to enable efficient as well as large-scale similarity search without significant fidelity loss.
    \item We demonstrate through experiments that ULTRA sets a new performance benchmark, consistently achieving precision rates exceeding 90\%. It also significantly outperforms existing non-adaptive semantic and lexical baselines for Urdu text based content recommender systems.
\end{itemize}

\begin{figure*}[htbp]
\centering
\includegraphics[width=1\textwidth, trim=20 40 20 50, clip]{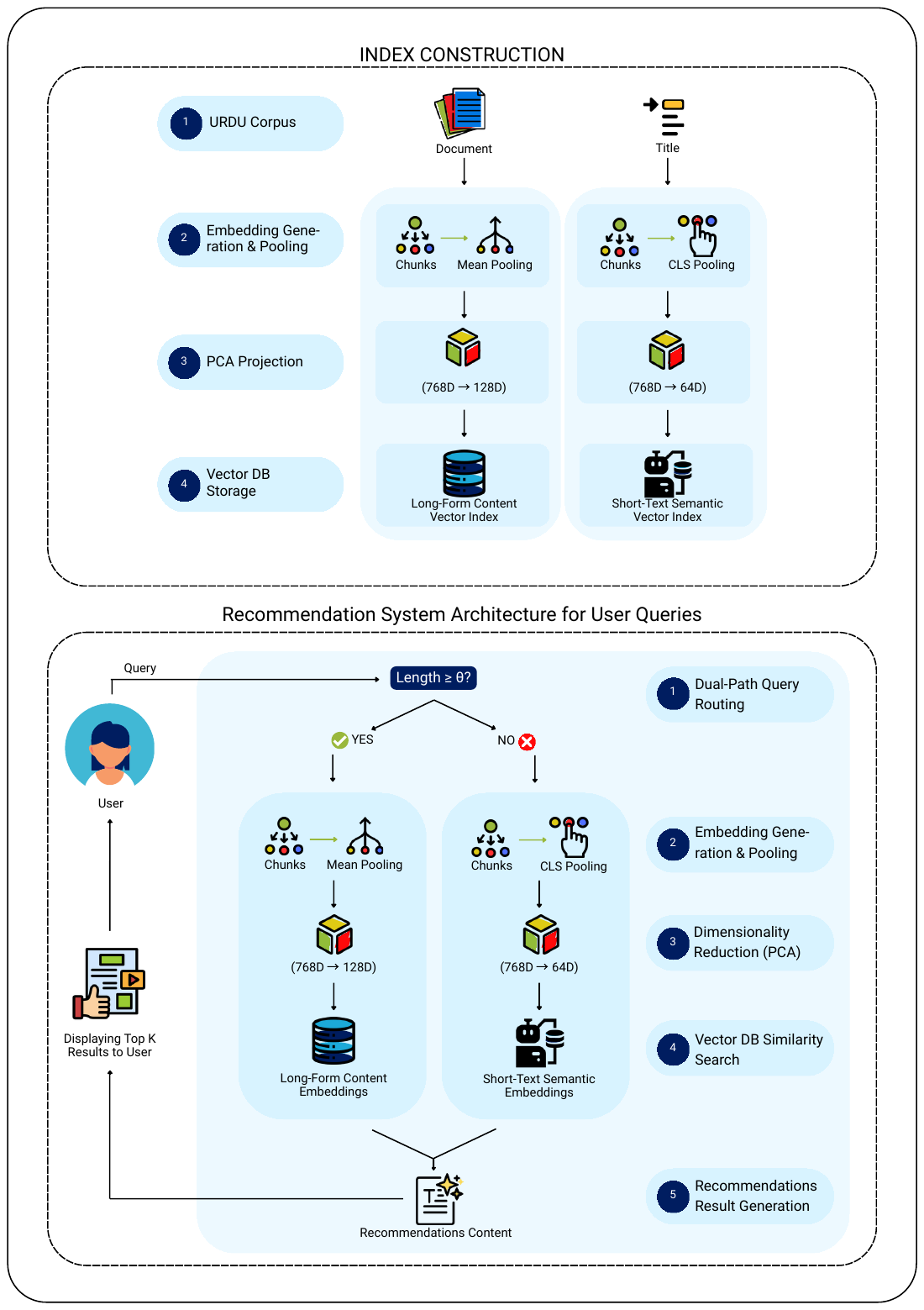}
\caption{Architecture diagram of proposed ULTRA framework}
\label{fig:Art}
\end{figure*}

\section{Related Work}
\label{sec_related_work}

Text-based content recommendation systems have evolved significantly, leveraging advancements in natural language processing (NLP) to provide personalized experiences across various domains, including news, articles, social media, and e-commerce. Early systems primarily focused on content-based filtering, where recommendations are generated based on the similarity between item features and user preferences, often using vector space models \cite{1}. For instance, a foundational survey on recommender systems highlighted the transition from simple keyword-based methods to more sophisticated hybrid approaches that incorporate user behavior and contextual data \cite{30}. These systems aim to address information overload by curating relevant text content, but they often struggle with semantic depth, especially in diverse linguistic contexts \cite{2}.

 In high-resource languages like English, content recommendation has benefited from large-scale datasets and advanced models. A comprehensive review of text-based recommendation systems emphasized the role of feature selection techniques, such as word embeddings, in improving accuracy over traditional methods \cite{14}. This survey analyzed over 100 studies, concluding that hybrid models combining content-based and collaborative filtering yield better personalization \cite{31}. Another key work proposed a content-based book recommender using machine learning for text categorization, demonstrating how information extraction can enhance relevance in textual domains \cite{32}. However, these approaches often overlook scalability issues in real-time applications, where high-dimensional representations lead to computational bottlenecks \cite{33}.

 The integration of deep learning has marked a paradigm shift in text-based content recommendation. A survey on deep learning-based recommender systems categorized models into neural collaborative filtering, autoencoders, and convolutional networks, showing their superiority in capturing non-linear relationships in text data \cite{34}. More recently, foundation models have been adapted for recommendation, with a taxonomy covering feature-based, generative, and agentic paradigms \cite{35}. These models excel in handling sequential user interactions, but their application to text-heavy content requires careful embedding strategies \cite{36}.

 Transformer-based architectures have particularly revolutionized content recommendation by enabling better semantic understanding. The BERT-based Sequential Transformer (BST) model, for example, uses transformer layers to model user behavior sequences, achieving state-of-the-art performance on datasets like MovieLens \cite{37}. Transformers4Rec, an open-source library bridging NLP and recommender systems, facilitates the use of pre-trained transformers for session-based recommendations \cite{38}. In e-commerce, transformer models have been employed for substitute product recommendation, incorporating weakly supervised customer data to improve accuracy \cite{39}. Music recommendation systems have also adopted transformers to capture sequential patterns in user listening histories \cite{40}. A study on exploiting deep transformer models in review-based recommenders showed improved ratings prediction by processing textual reviews \cite{41}. Furthermore, a transformer-based architecture for collaborative filtering, MetaBERTTransformer4Rec, outperformed baselines in personalized settings \cite{42}. These works highlight the flexibility of transformers in text-based scenarios, but they often assume abundant resources, limiting applicability to diverse content types \cite{47}.

 Scalability and efficiency remain critical challenges in transformer-based systems. Techniques like dimensionality reduction via Principal Component Analysis (PCA) or UMAP have been explored to compress embeddings for large-scale vector searches \cite{22}. A survey on sequential recommendation systems discussed how transformers handle time-series data but emphasized the need for optimization in high-dimensional spaces \cite{43}. In real-time applications, such as at Scribd, transformer models have been deployed for content recommendation, balancing accuracy with latency \cite{44}.

Despite these advancements, low-resource languages pose unique hurdles for text-based content recommendations Low-resource settings, characterized by limited labeled data and linguistic tools, hinder the development of effective NLP models \cite{26}. A survey on Urdu language processing outlined challenges like morphological complexity and code-mixing, which affect semantic tasks \cite{25}. In multilingual contexts, recommender systems often underperform for low-resource languages due to biased training data \cite{45}. For instance, evaluations of large language models (LLMs) on low-resource languages like Bengali revealed gaps in zero-shot and few-shot prompting, relevant for recommendation personalization \cite{46}. Multimodal approaches have been proposed to augment text-based systems in low-resource scenarios, but they require integrated frameworks \cite{47}.

 Focusing on Urdu, a low-resource language with over 230 million speakers, text-based content recommendation remains underdeveloped. Existing research has primarily explored foundational techniques, often adapted from high-resource paradigms. Early systems relied on traditional lexical methods, such as TF-IDF with cosine similarity, which are limited by surface-level keyword overlap \cite{14,15}. A review of text-based recommendation systems discussed these challenges in low-resource languages like Urdu, advocating for advanced NLP to bridge semantic gaps \cite{14}. Acknowledging this, subsequent work integrated pre-trained transformer models like BERT to generate semantic embeddings \cite{6}. For Urdu content, a content-based model using UrduHack for preprocessing and BERT embeddings achieved higher accuracy in personalized recommendations \cite{16}.

 Further innovations have enriched content representation through entity-based techniques. A foundational contribution in this direction was the development of an Urdu Wikification pipeline, which proposed a novel approach for linking named entities within Urdu text to their corresponding pages in Wikidata \cite{17}. This work was pivotal for multiple reasons. First, it created and made publicly available valuable language-specific resources, including the SKEL dataset (550 annotated news titles) and the larger SKRS dataset (16,738 articles with user interaction data), which provided a critical infrastructure for subsequent research. Second, it demonstrated a practical methodology for constructing a sub-knowledge graph (8,439 entities and 23,080 relationships) from these linked entities. By applying the TransE algorithm to this graph to generate knowledge graph embeddings, the authors created a rich, structured semantic layer that could be integrated into a news recommendation model. The reported final accuracy of 60.8\% for this hybrid RNN-based recommender established a new, knowledge-enhanced benchmark over purely lexical baselines for Urdu, showcasing the potential of leveraging structured external knowledge.

 Building directly upon this wikification foundation, the later work SEEUNRS (Semantically Enriched Entity-Based Urdu News Recommendation System) refined and scaled the entity-centric approach \cite{24}. This framework utilized an expanded dataset of 23,250 Urdu news articles to exploit the "hidden semantic features" captured through entity and knowledge graph embeddings. The core advancement of SEEUNRS was its deeper integration of this entity knowledge to drive personalized recommendations. Rather than treating entities merely as tags, the system used their embeddings to model nuanced semantic relationships between users, articles, and concepts. This enabled the framework to outperform traditional lexical methods (e.g., TF-IDF) in precision, demonstrating that semantic enrichment through entities could more effectively address the vocabulary mismatch problem and capture user interest in low-resource settings.

 Limitations of prior Urdu work include: (1) reliance on global, single-representation matching (no distinction between headline vs. full-text semantics); (2) no exploration of multiple pooling strategies or explicit evaluation of pooling choices; (3) absence of dimensionality-reduction and vector-storage engineering (no vector DB), which constrains large-scale retrieval and latency; and (4) no adaptive handling of query granularity (short vs. long queries are treated uniformly), hence not accommodating multiple use cases of recommendation results \cite{27}. 

 Our ULTRA framework addresses these gaps by introducing a dual-pathway system with query routing, optimized pooling, dimensionality reduction, and vector databases. This advances Urdu text-based content recommendation, generalizing beyond news to handle diverse textual corpora efficiently. Table \ref{tab:comp} outlines the key limitations of prior work in Urdu text-based language processing systems and illustrates how our proposed ULTRA framework addresses them.

\begin{sidewaystable*}[ph!]
\FloatBarrier
\caption{Comparative analysis of foundational Urdu NLP systems and their contributions to text-based content recommendation}
\label{tab:comp}
\bigskip
    \centering\small \setlength\tabcolsep{4 pt}
    \setlength{\extrarowheight}{12pt}
        \hspace*{-.2cm}
\begin{tabular}{p{1.6cm}p{1.5cm}p{3cm}p{2.5cm}p{3cm}p{5cm}>{\raggedright\arraybackslash}p{5cm}}
\toprule
\textbf{Ref.} & \textbf{Yr.} & \textbf{Model/Technique} & \textbf{Dataset} & \textbf{Paradigm}& \textbf{Key Contributions} & \textbf{Identified Limitations} \\ 
\cline{1-7}

\cite{14}& 2021 & Literature Review / Survey & N/A (Review Paper) & Survey of Text-based RS & Comprehensive review of methods; Identified core challenges for low-resource languages like Urdu. & No implementation; highlights gaps but provides no solution. \\

\cite{16}& 2022 & TF-IDF + Cosine Similarity, BERT Embeddings & Custom Urdu News (1,160 articles) & Content-based Filtering & Introduced semantic BERT embeddings for Urdu news, moving beyond pure lexical matching. & Limited dataset; single-pipeline design; no adaptation for different query types; relies partly on lexical matching. \\

\cite{17}& 2022 & Entity Linking + RNN Hybrid with TransE & SKRS Dataset (16,738 articles) & Knowledge Graph-based Hybrid & Pioneered Urdu Wikification; used entity knowledge graph to enrich semantics for recommendation. & Moderate accuracy (60.8\%); complex pipeline; performance tied to entity presence; no query adaptation. \\

\cite{24}& 2024 & Entity-Enriched Framework (SEEUNRS) & Expanded Dataset (23,250 articles) & Entity-Based Semantic Recommendation & Advanced entity-based matching, exploiting hidden semantic features for improved precision. & Heavy reliance on entity recognition; no query-adaptive routing; limited scalability for real-time systems. \\

\cite{50}& 2023 & Deep Learning for Urdu News Classification & Urdu News 1M Dataset (1M articles) & Text Classification & Large-scale Urdu news classification using transformers; demonstrated scalability. & Classification-focused; no personalization or recommendation components; single-representation approach. \\

\cite{51}& 2021& RoBERTa-based Pre-trained Language Model& Urdu news corpus& Language Model Pre-training& Earliest public RoBERTa-based model for Urdu; handles right-to-left Nastaliq script, morphology, and semantics; foundational for Urdu semantic NLP tasks.& Focused on language modeling only; no recommendation-specific fine-tuning or retrieval optimization. \\

\cite{52}& 2023 & Multilingual Sentence-BERT for Indic Languages & Synthetic NLI/STS& Cross-lingual Embeddings & Extended sentence transformers to low-resource Indic languages; improved semantic similarity.&General-purpose embeddings; not optimized for recommendation tasks; no query routing mechanism. \\

\cite{53}& 2026& U-RR²: Two-stage framework integrating TF-IDF, dense embeddings, and SVMrank re-ranking with feature ensemble & CURE, ROSHNI, UIR\_21 benchmarks & Two-Stage Dense Retrieval \& Re-ranking & Proposes a hybrid two-stage Urdu document retrieval framework with large-scale word embeddings, achieving state-of-the-art performance on Urdu IR benchmarks (MAP 0.78–0.81, P@10 0.79–0.82, $F_1$
 up to 0.90).& Designed for ad-hoc document retrieval rather than personalized semantic recommendation; lacks query-adaptive routing, granularity differentiation, recommendation-specific optimizations (pooling strategies, dimensionality reduction), and scalable real-time personalization mechanisms.\\
 
\textbf{ULTRA}& \textbf{2026} & \textbf{Dual-Embedding Transformer with Query-Adaptive Routing} & \textbf{Large-Scale Corpus (\textasciitilde112k articles)} & \textbf{Adaptive Semantic Recommendation} & \textbf{Novel dual-pathway architecture with length-threshold routing; optimized pooling \& PCA; scalable vector DB integration.} & \textbf{Addresses prior gaps: query-length variability, single-representation limitation, and scalability.} \\

\bottomrule
\end{tabular}\hspace*{-0.1cm}
\end{sidewaystable*}

\section{Proposed Methodology}
\label{sec_proposed_methodology}
In this research, we propose a methodology to develop a robust Urdu news recommendation system using semantic search powered by transformer-based embeddings. The architecture of the proposed system is illustrated in Figure \ref{fig:Art}, and its modules are described below.

\subsection{Problem Formulation}
\label{sub_sec_problemFormulation}

Considering a user query in Urdu and a collection of Urdu news stories, the task is to find the top $k$ semantically relevant news stories. We state this as a semantic search problem in high-dimensional vector space.

Suppose the article corpus is a set $D = \{d_1, d_2, \ldots, d_m\}$ where each article $d_i$ is a pair: $d_i = (h_i, c_i)$ where $h_i$ is the headline and $c_i$ is the content/complete news article that is the subject of that headline.

The query made by a user is represented by $q \in \mathcal{Q}$ with the space of all possible Urdu text queries denoted by $\mathcal{Q}$. The length of a query is provided by the function $\ell(q)$.

The main activity of the system is to access a collection of articles that are semantically related to a query of a user. We represent this by a binary relevance function, which we denote as $\text{rel}: \mathcal{Q} \times D \rightarrow \{0, 1\}$. Given a query $q$ and an article $d_i$, a human evaluator assigns:
\[
\text{rel}(q, d_i) = 
\begin{cases} 
1, & \text{if article } d_i \text{ is relevant to query } q, \\
0, & \text{otherwise.}
\end{cases}
\]
A True Positive (TP) is a relevant article that has been returned in response to a query.

The system uses cosine similarity in learned embedding space as an effective proxy to estimate semantic relevance operationally. It returns the first $k$ articles, denoted $R_q^k$, in terms of this similarity score:
\begin{equation}
R_q^k = \underset{d_i \in D, |R|=k}{\operatorname{argmax}} \cos(\mathbf{v}_q, \mathbf{v}_i).
\label{eq:retrieval}
\end{equation}
Precision is the key performance indicator. For a single query $q$, the precision at $k$ (Precision@k) is the proportion of retrieved articles that are relevant:
\begin{equation}
\text{Precision@k}(q) = \frac{1}{k} \sum_{d_i \in R_q^k} \text{rel}(q, d_i) = \frac{|\text{TP}_q|}{|\text{TP}_q| + |\text{FP}_q|},
\end{equation}
where $\text{TP}_q$ is the set of True Positives and $\text{FP}_q$ is the set of False Positives (retrieved but irrelevant articles) for query $q$.

To evaluate the system's overall effectiveness, we compute the mean precision across a test set of $N$ queries:
\begin{equation}
\text{Mean Precision} = \frac{1}{N} \sum_{q \in Q_{\text{test}}} \text{Precision@k}(q).
\label{eq:mean_precision}
\end{equation}

\subsubsection{Embedding generation}
Each textual element, be it headline or news article (named as content in our corpus), is mapped to a dense vector representation (embedding) using a pre-trained Urdu transformer model $\mathcal{M}$. The model generates a contextual embedding for each token. A pooling function $ \mathcal{P} $ is applied to aggregate these token embeddings into a single, fixed-dimensional sentence embedding vector.
\begin{align}
\mathbf{v}_q &= \mathcal{P}(\mathcal{M}(q)) \in \mathbb{R}^{768}, \\
\mathbf{v}_{h_i} &= \mathcal{P}(\mathcal{M}(h_i)) \in \mathbb{R}^{768}, \\
\mathbf{v}_{c_i} &= \mathcal{P}(\mathcal{M}(c_i)) \in \mathbb{R}^{768}.
\end{align}
For the content $ c_i $, if its token length exceeds the model's maximum sequence limit $ L_{\max} $, it is segmented into overlapping chunks $\{c_i^{(1)}, \ldots, c_i^{(n)}\}$. The final content embedding is the average of the individual chunk embeddings:
\begin{equation}
\mathbf{v}_{c_i} = \frac{1}{n} \sum_{j=1}^{n} \mathcal{P}(\mathcal{M}(c_i^{(j)})).
\end{equation}

\subsubsection{Query-adaptive routing}
\label{sub_sub_sec_adaptive_routing}
To align the granularity of the query with the appropriate semantic representation, a threshold $ \theta $ on query length governs the retrieval pathway:
\[
\text{Retrieval Space} = 
\begin{cases}
\{\mathbf{v}_{h_i}\}_{i=1}^m, & \ell(q) < \theta \\
\{\mathbf{v}_{c_i}\}_{i=1}^m, & \ell(q) \geq \theta 
\end{cases}
\ \ {\small\begin{array}{l}
\text{(Short Query Route)} \\
\text{(Long Query Route)}
\end{array}}
\]
The selection of the threshold \( \theta \) is determined empirically based on the average length of titles or headings within the content corpus of the recommender system. In our case, corpus comprised of news articles, we approximated the headline length, given that headline embeddings (\( \mathbf{v}_{h_i} \)) are utilized for semantic comparison against short queries, and adopted this value as \( \theta \) to delineate short from long queries. The underlying intuition for distinguishing between long and short queries and routing them to the appropriate pathway is already elaborated in Section~\ref{sec_novelty}.

\subsubsection{Dimensionality reduction for efficiency}
To optimize large-scale retrieval, a dimensionality reduction function $ \mathcal{R}: \mathbb{R}^{768} \rightarrow \mathbb{R}^{d'} $, where $ d' \ll 768 $, is applied to the pooled embeddings. For a given pathway, the reduced vectors are:
\begin{equation}
\mathbf{v}'_{*} = \mathcal{R}(\mathbf{v}_{*}), \quad * \in \{q, h_i, c_i\}.
\end{equation}
This step preserves essential semantic relationships while significantly reducing the computational cost of similarity search.

\subsubsection{Similarity computation and ranking}
The semantic similarity between the query $ q $ and an article $ d_i $ is computed as the cosine similarity between their corresponding embeddings in the chosen (and potentially reduced) vector space. Let $\mathbf{v}'_q$ be the query embedding and $\mathbf{v}'_i$ be the target article embedding (either $\mathbf{v}'_{h_i}$ or $\mathbf{v}'_{c_i}$). The similarity score is:
\begin{equation}
\text{sim}(q, d_i) = \cos(\mathbf{v}'_q, \mathbf{v}'_i) = \frac{\mathbf{v}'_q \cdot \mathbf{v}'_i}{\|\mathbf{v}'_q\| \|\mathbf{v}'_i\|}.
\end{equation}
This cosine score serves as the operational retrieval metric. The final ranked list $R_q^k$ is obtained by sorting all articles in descending order by $\text{sim}(q, d_i)$ and selecting the top $k$. It is important to note that while this geometric measure is an effective proxy for semantic relatedness, the final assessment of recommendation quality is based on human-evaluated relevance, as defined in Section \ref{sub_sec_problemFormulation}. This system aims to maximize the average precision by optimizing the underlying embedding generation, routing, and dimensionality reduction modules to generate $R_q^k$ that is as close as possible to human relevance judgments.

This formulation gives a strict mathematical basis to the further development and analysis of the components of the ULTRA framework, such as the selection of $P$ (pooling strategies), the determination of $\theta$ (routing threshold), the selection of $R$ (dimensionality reduction technique) and optimization of $d$ (reduced dimension). It is important to note that this problem formulation is universal and can be used with any recommendation system, but in our implementation, we use news articles because of the availability of the largest dataset of Urdu corpus to evaluate. However, the framework is applicable to a broad range of recommendation systems, which include digital libraries, news recommenders, product recommenders, and blog page recommenders.

\subsection{Dataset}
\label{sub_sec_ds_Preprocessing}

\begin{figure}[h]
  \centering
  \includegraphics[width=1\linewidth, trim=0 0 0 0, clip]{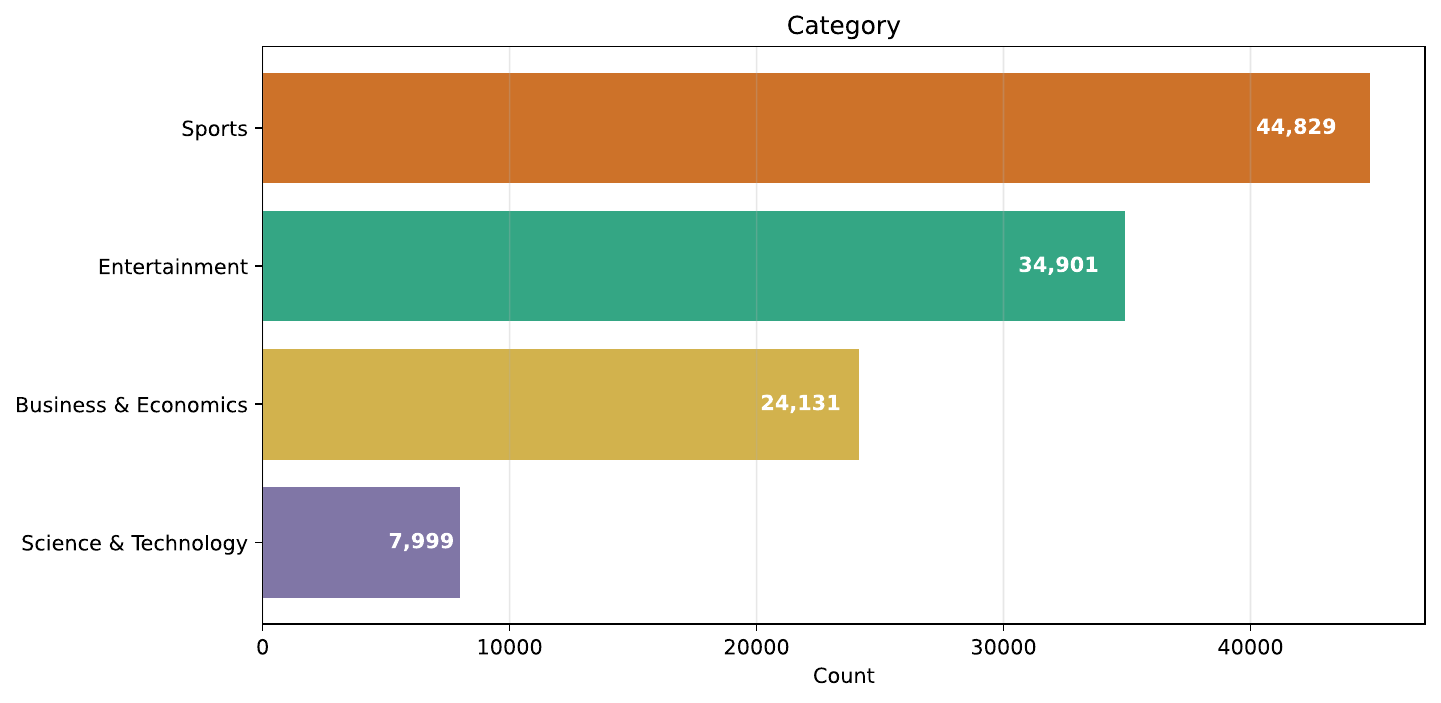}
  \caption{Graphical representation of categories by frequency of Urdu news dataset \cite{50}}
  \label{fig:cat}
\end{figure}

The source of our dataset is Kaggle \cite{50}. Our experiments used pre-processing of about a corpus of 112k articles. The data set includes 4 distinct categories i.e., Science \& Technology, Business \& Economics, Entertainment \& Sports. An overview of the articles that are categorized into various categories in is presented in Figure \ref{fig:cat}.

\subsection{Dataset Preprocessing
}
\label{sub_sec:preprocessing}

\begin{figure}[h]
  \centering
  \includegraphics[width=1\linewidth,  trim=80 830 80 200, clip]{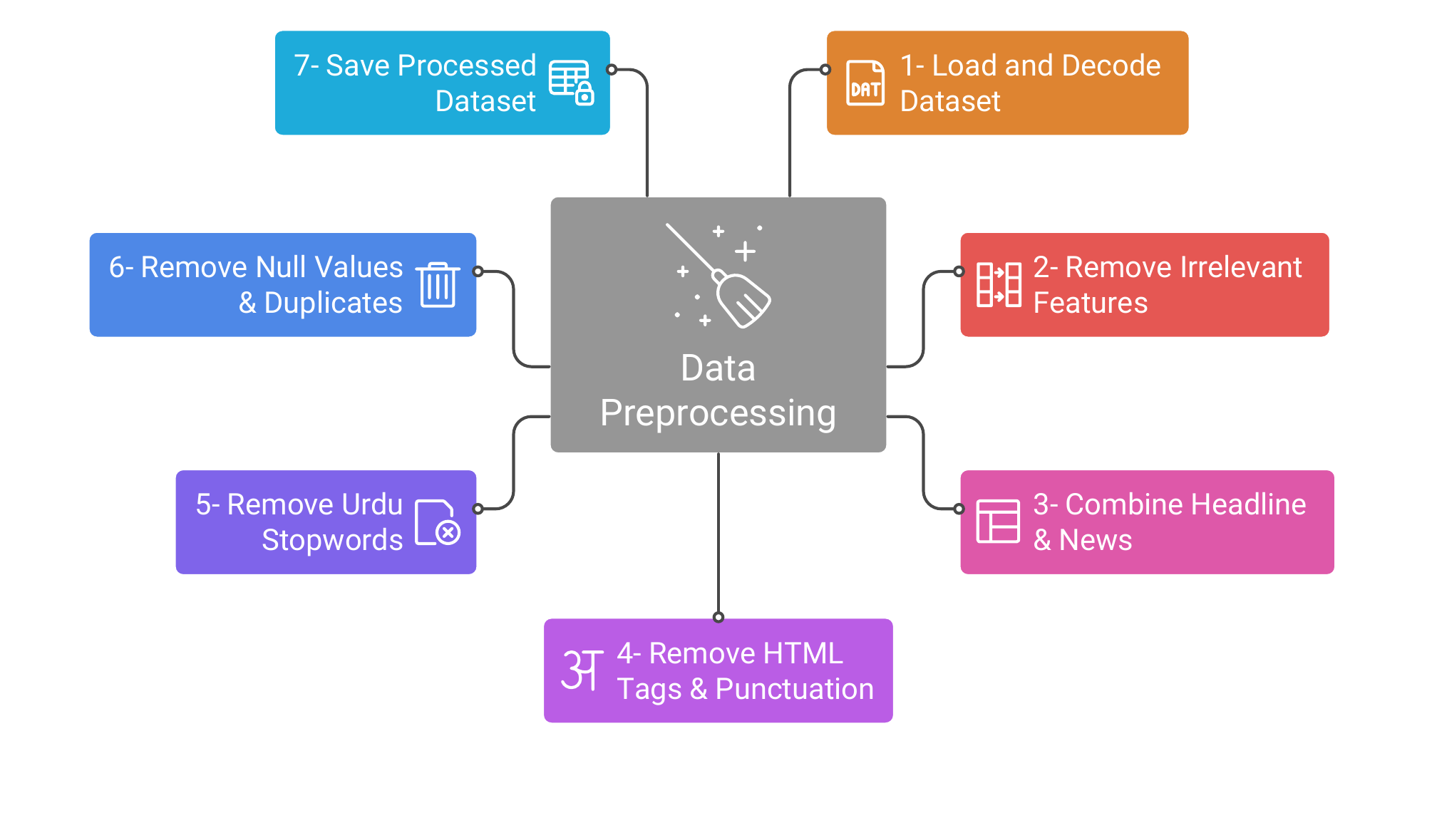}
  \caption{Urdu news dataset pre-processing steps}
  \label{fig:pre}
\end{figure}

The raw Urdu news data was subjected to a systematic pre-processing pipeline to guarantee quality and consistency to be used in further analysis. This was done through five major steps: (1) Dataset Loading and Encoding Normalization, during which the loading of the first corpus was conducted, and encoding problems were addressed to fix the representation of Urdu characters. (2) Metadata Pruning and Composite Text Construction, which entails eliminating non-informative metadata (e.g., author, date stamps not used in this study) and concatenation of article headlines with respective article text to create a single content field. (3) Text Cleaning and Normalization, which removed HTML tags, punctuations, special characters, and unnecessary whitespace to standardize the textual data. (4) Urdu Stop-Word Elimination, in which a list of high-frequency, low-information Urdu words was weeded to remove noise. (5) Invalid and Redundant Record Filtering, in which records that contained a null value, exact duplicate or those that had a text length less than a specified threshold were deleted to complete the dataset. Visual representation of the pre-processing steps are depicted in Figure \ref{fig:pre}.
 
\subsubsection{Dataset loading and encoding correction
}
The Urdu text datasets usually have a problem of inconsistencies in character encoding when handled using the normal file reading procedures. The problem is caused by the differences in file encoding formats and text processing that is platform-dependent. In order to maintain the integrity of the Nastaliq script with its complex combinations of characters and diacritical marks, the dataset was first loaded with the encoding scheme of the \texttt{ISO-8859-1}.

To identify and fix mis-encoded characters, a custom decoding function was used. This feature works by identifying the mismatch in the encoding based on the character pattern, re-encoding troublesome text snippets in the  \texttt{ISO-8859-1} to \texttt{UTF-8}, and comparing the corrected text with Urdu Unicode ranges.

This methodical correction guaranteed the correct representation of all Urdu characters, that include special diacritics (such as \textit{zabar}, \textit{zer}, and \textit{pesh}) and punctuation signs characteristic of Urdu script. Encoding normalization was also done to all textual columns: \texttt{Headline}, \texttt{News Text}, \texttt{Category}, and \texttt{Source}. This was a crucial step to ensure that no information is lost in the later stages of processing.

\subsubsection{Removal of irrelevant metadata and composite content creation}

News datasets are usually characterized by a range of metadata fields that, although they may be useful in certain analysis, do not help to understand the semantics in text classification and embedding tasks. Non-informative metadata columns were removed systematically to reduce the dimensions and possible noise factors. These were: \texttt{Index}: Sequential identifiers that are not semantically meaningful, \texttt{Date}: Publication dates (not used in this text-oriented study), \texttt{URL}: Web addresses with platform-specific patterns, \texttt{Source}: Publication sources (not used in this text-oriented study), and \texttt{News Length}: Derived statistics, not the raw content.

After metadata pruning, a single textual representation was developed to achieve in-depth article semantics. The  \texttt{Headline} and \texttt{News Text} columns were joined together with a delimited joining technique to create a new column called content. This field is a composite which maintains the terse, information-rich headlines. It takes into account the contextual information of the article bodies in more detail, as well as builds a single input stream to embed models, which makes pipeline architecture simpler.

\begin{figure}[h]
  \centering
  \includegraphics[width=1\linewidth, trim=20 190 20 40, clip]{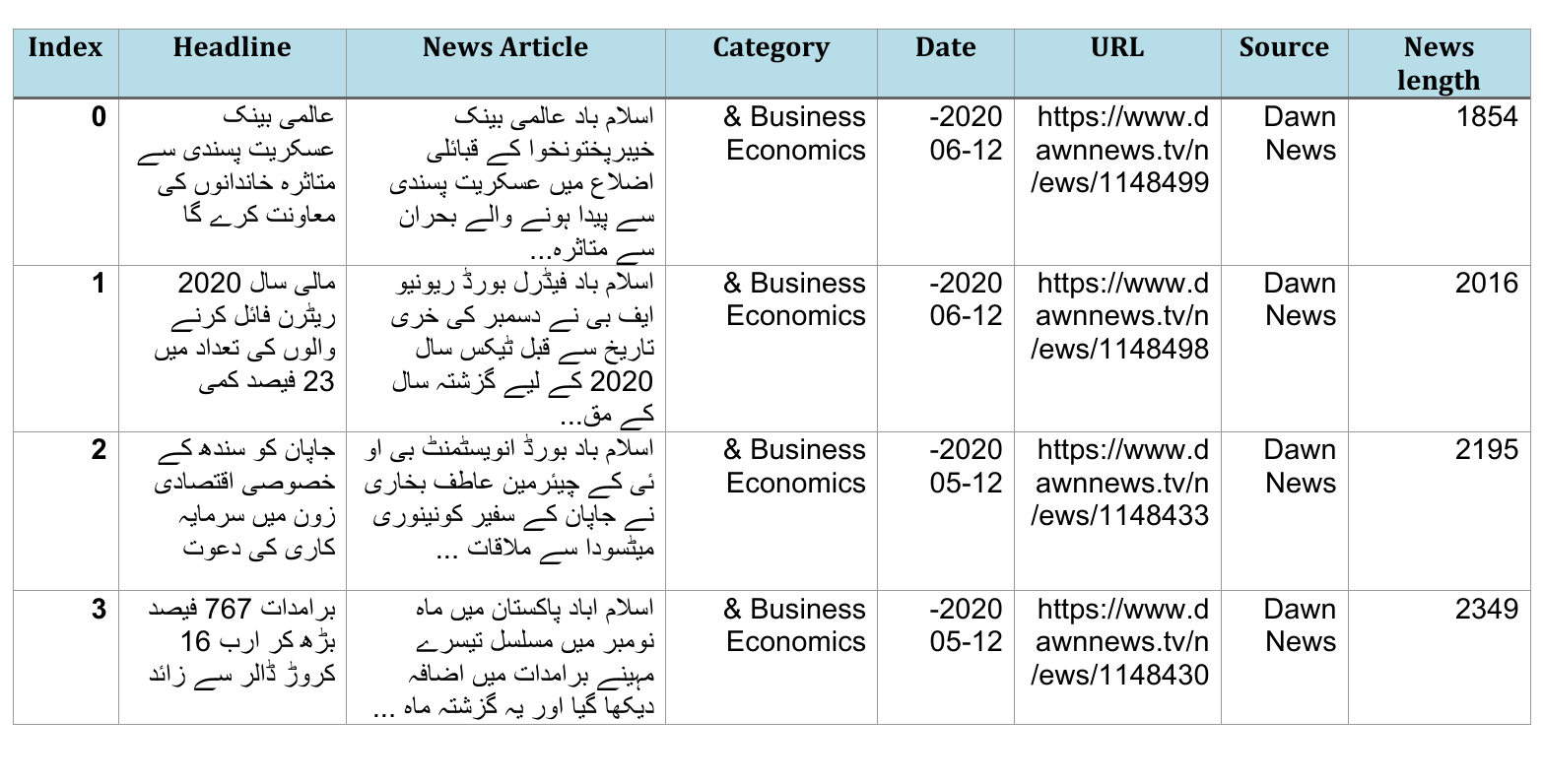}
  \caption{Dataset before removal of irrelevant metadata}
  \label{fig:T1}
\end{figure}

\begin{figure}[h]
  \centering
  \includegraphics[width=1\linewidth, trim=60 242 40 60, clip]{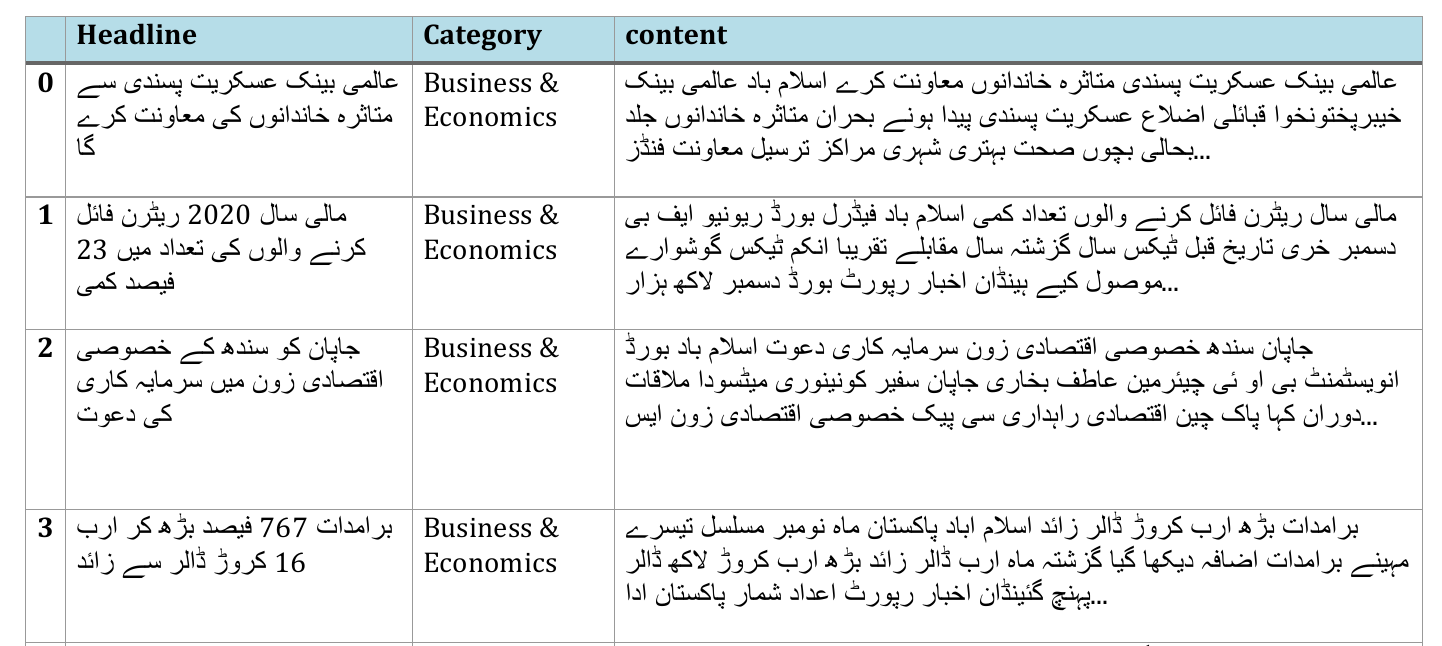}
  \caption{Dataset after removal of irrelevant metadata}
  \label{fig:T2}
\end{figure}

The change is such that the models are provided with topic-signaling role of headlines as well as the elaboration of full articles. Figure \ref{fig:T1} and Figure \ref{fig:T2} shows the structure of the dataset prior to this transformation step and after it respectively.

\subsubsection{Text cleaning and normalization}

Web sources of raw text frequently include structural artifacts, remnant formatting and irregular use of characters which may have a negative impact on downstream processing. In order to produce a clean, standardized corpus, a multi-layered text normalization pipeline was used:

\textit{a)- HTML and formatting tag removal:}
The content scraped by the Web often contains HTML tags, CSS classes and JavaScript fragments. These were stripped with a mix of regular expression and HTML parsing libraries so as to remove them completely leaving behind textual content.

\textit{b)- Non-Urdu character elimination:}
Since it was about the analysis of Urdu characters, irrelevant sets of characters were eliminated: English alphabets (A-Z, a-z), Arabic numerals (0-9), mixed-language characters and signs, platform specific items (e.g., emoticons, emojis which are not applicable in the news context).

\textit{c)- Punctuation normalization:}
Systematic removal of both Urdu-specific and generic punctuation marks was done to form word-level tokens with no interference of boundaries: generic punctuation: `.` (period), `,` (comma), `;` (semicolon), `:` (colon), quotation marks, brackets of any type, Urdu comma, Urdu semicolon, Urdu question mark.

\textit{d)- Whitespace standardization:}
Several successive whitespace characters (spaces, tabs, newlines) were contracted to one, and initial and final whitespace was riddled out. This normalization guarantees that there is a consistency in the token boundaries when further processing is done. 

The effect of these cleaning processes is a uniform Urdu text corpus in which semantic information is maintained and structural noise is reduced. This clean text is ideal input to any traditional NLP pipeline as well as modern embedding models.

\subsubsection{Urdu stop-word elimination}

\begin{figure}[h]
  \centering
  \includegraphics[width=1\linewidth, trim=70 580 70 70, clip]{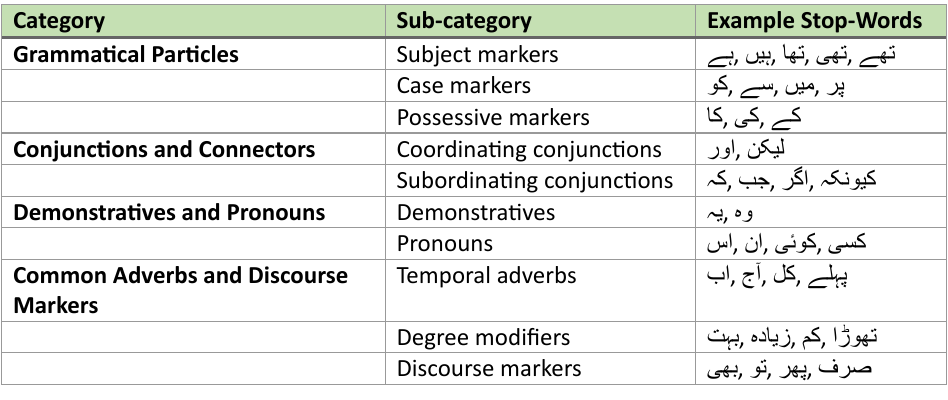}
  \caption{Categories of Urdu functional words included in the stop-word list}
  \label{fig:Stop_Word}
\end{figure}

The text representations can be dominated by stop words that do not provide meaningful information. In the case of Urdu, where the grammatical particles and connectors are abundant, the removal of stop-words is especially effective. A list of 127 Urdu stop words was created manually as a result of linguistic analysis and frequency distribution research of the corpus. A stop-word list contains multiple categories of function words:

\textit{a)- Grammatical particles:}
Words that primarily serve grammatical functions rather than conveying content as shown in Figure~\ref{fig:Stop_Word}.

\textit{b)- Conjunctions and connectors:}
Words that connect phrases but have limited standalone meaning. Figure~\ref{fig:Stop_Word} demonstrates some coordinating and subordinating conjunctions.

\textit{c)- Demonstratives and pronouns:}
Common pronouns and demonstratives that appear frequently but lack specific content. Figure~\ref{fig:Stop_Word} gives some examples of demonstratives and pronouns.

\textit{d)- Common adverbs and discourse markers:}
Frequently occurring adverbs and discourse particles as in  Figure~\ref{fig:Stop_Word}.

\begin{figure*}[h]
  \centering
  \includegraphics[width=0.7\linewidth, trim=20 60 20 20, clip]{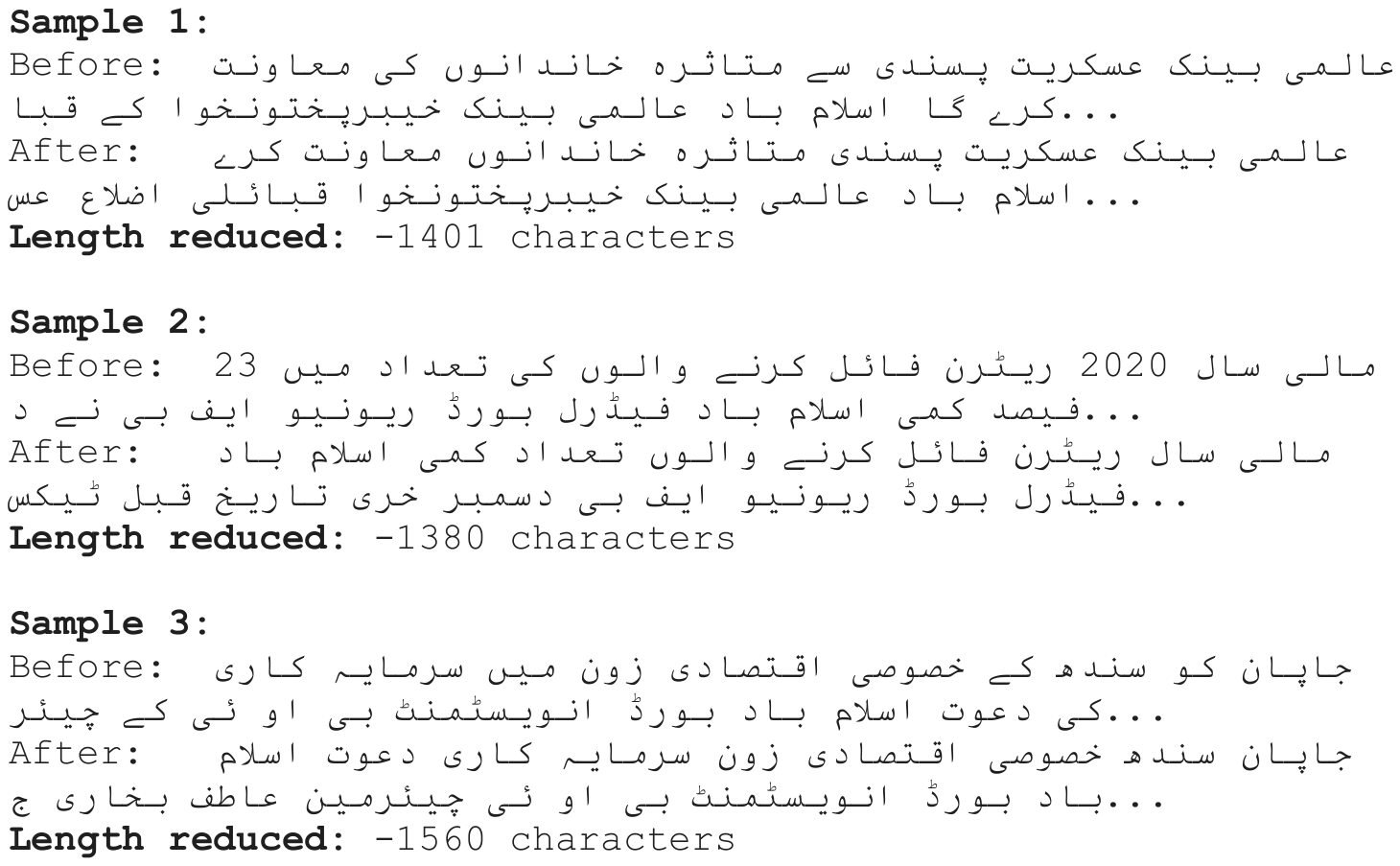}
  \caption{Common adverbs and discourse markers in Urdu language}
  \label{fig:SWR}
\end{figure*}

The removal procedure followed the cleaning of the text but preceded any tokenization or feature extraction. This way, we can recognize stop words in their normal form thereby enhancing accuracy in removal. The effectiveness of Urdu stop-word removal can be seen in the three examples presented in Figure~\ref{fig:SWR}. In every example, there is the original (before) text, the text after filtering (after), and the number of characters reduced, and this shows that there is a great deal of compression yet the meaning has not been lost.

\begin{figure*}[h]
  \centering
  \includegraphics[width=0.9\linewidth, trim=80 300 80 70, clip]{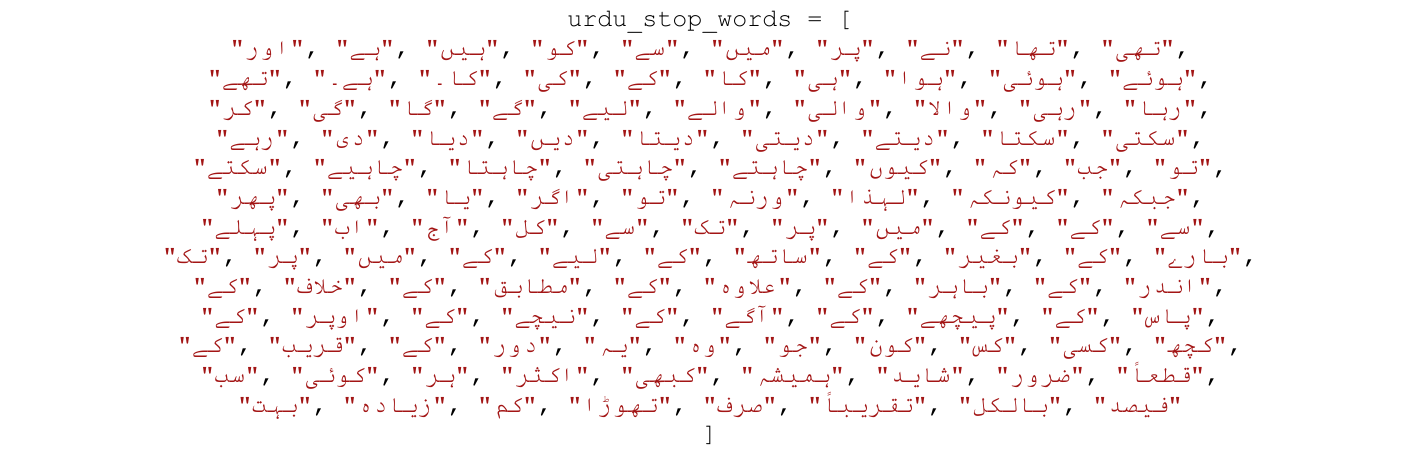}
  \caption{Common stop words in urdu language}
  \label{fig:Sw}
\end{figure*}

Stop-word elimination decreases the size of the corpus by approximately 33.5\% quantitatively, indicating the high rate of such functional words within the Urdu news. This automatic shrinkage reduces the computational requirements and can improve the performance of the model by focusing on content carrying terms. Figure~\ref{fig:Sw} shows 127 stop words that were identified and eliminated.

\subsubsection{Removal of invalid and redundant entries}

The assurance of data quality is also essential to creating trustworthy machine learning models. The pre-processing pipeline removes problematic records based on the following  criteria:

\textit{a)- Detection and Removal of Null Values:}
Records that lack essential information in significant columns are deleted. Our targets are the \texttt{content} column (Empty or null text fields,) and the \texttt{Category} column (unlabeled or vague categories) and critical metadata (absent values). This ensures that all training examples are fully featured and labeled.

\textit{b)- Identifying Duplicate Content:}
Duplicate articles (because of republication, syndication, or crawl artifacts) can create bias, as they over-represent part of the content. This move keeps diversity in the datasets and eliminates redundancy.

\textit{c)- Minimum length filtering:}
Text entries that are less than 15 characters do not necessarily have enough semantic meaning. They can be fragments of articles, text of navigation, or placeholders mistakenly captured. This is done to eliminate such entries to retain only substantive content.

\subsubsection{Post-Processing Dataset Structure}
\label{sub_sec:postprocessing_structure}

After the entire pre-processing process, the dataset is organized in the most suitable way to be used in downstream activities. It is encoded in a table format and utilizes UTF-8 encoding, which is compatible with Urdu text processing libraries and frameworks. This is a clean labeled dataset that is used as a starting point to all further operations presented in the subsequent sections.

A combination of these filtering operations results in a clean, uniform dataset. Table~\ref{tab:preprocessing_stats} shows the quantitative effects of the pre-processing on the dataset properties.

\begin{table}[ht]
  \centering
  \begin{tabular}{@{}p{2cm}p{4cm}p{1.4cm}@{}}
  
    \toprule
    \textbf{Metric} & \textbf{Description} & \textbf{After Preprocessing} \\
    \midrule
    Records        & Total number of articles & 111,853 \\[2pt]
    Article Length (avg.)& Average characters per article & 986.3\\[2pt]
 Headline Length (avg.)& Average characters per headline&52.31\\
    Words          & Total words processed & 30,639,468\\[2pt]
    Stop Words     & High-frequency tokens removed & 10,257,229\\[2pt]
    Removal Rate   & Percentage of stop words removed & 33.5\% \\[2pt]
    Categories     & Unique topic categories & 5 \\[2pt]
    Sources        & Distinct publication sources & 6 \\[2pt]
    Duplicates     & Redundant articles removed & 8\\
    \bottomrule
   
  \end{tabular}
   \caption{Dataset statistics after preprocessing}
   \label{tab:preprocessing_stats}
\end{table}

\subsection{Proposed Methodology}
\label{sub_sec:methodology}

The ULTRA framework implements a semantic search-based recommendation system tailored for Urdu news. As formalized in Section \ref{sub_sec_problemFormulation}, the core challenge is to retrieve semantically relevant articles from a corpus of over 112k Urdu news items based on user queries of varying lengths and semantic granularity. By "query granularity," we refer to the level of detail and specificity in a user's search input, ranging from concise, keyword-like queries, to elaborate, descriptive sentences. However, in our framework, news articles or full-length documents predominantly function as long queries, targeted at the use case where recommendations are generated for content relevant to the article the user is currently viewing.

Our approach addresses this challenge through three interconnected components: (1) adaptive embedding generation that aligns with query granularity, (2) efficient dimensionality reduction for scalable retrieval, and (3) an optimized dual-path retrieval architecture. The entire methodology is structured to trade semantic accuracy and computational efficiency, to enable real-time recommendations with large-scale deployment.

\subsection{Adaptive Embedding Generation}

The process of semantic retrieval starts with the creation of high quality vector representations of both queries and documents that represent the semantics of the query and document. We used a pre-trained Urdu transformer model, \texttt{urduhack/roberta-urdu-small}, that generates 768-dimensional contextualized embeddings for input tokens \cite{51}. In order to map these token embeddings into a fixed-dimensional representation, for an entire sequence of text, we use a pooling strategy. 

\subsubsection{Pooling Strategy Analysis}
We tested and analyzed three types of pooling, which are mean pooling, max pooling, and CLS pooling. Mean pooling averages all token embeddings, which is a centroid of semantic information. Max pooling extracts the maximum value across each dimension, potentially emphasizing salient features. CLS pooling utilizes the special classification token's embedding, which is trained during pre-training to capture sentence-level semantics.

Initial evaluation revealed that the optimal pooling strategy depends on the nature of the text being encoded. For long-form article content, mean pooling demonstrated superior performance in capturing distributed semantic context. For concise headlines, CLS pooling proved most effective at preserving focused topical meaning. This observation informed the development of our dual-path embedding architecture.

\subsubsection{Query-Adaptive Dual-Path Architecture}
A key insight driving our methodology is the semantic granularity mismatch between user queries and article representations. Short, focused queries as in Figure~\ref{fig:Q1}
typically express a specific topic or intent in few words, making them semantically more aligned with article headlines or titles of documents, which are themselves concise summaries, than with the distributed, detailed content of full articles. To address this, we implemented a query-adaptive routing mechanism based on character length, which dynamically selects the most appropriate semantic space for comparison.

\begin{figure}[h]
  \centering
  \includegraphics[width=1\linewidth, trim=0 0 0 0, clip]{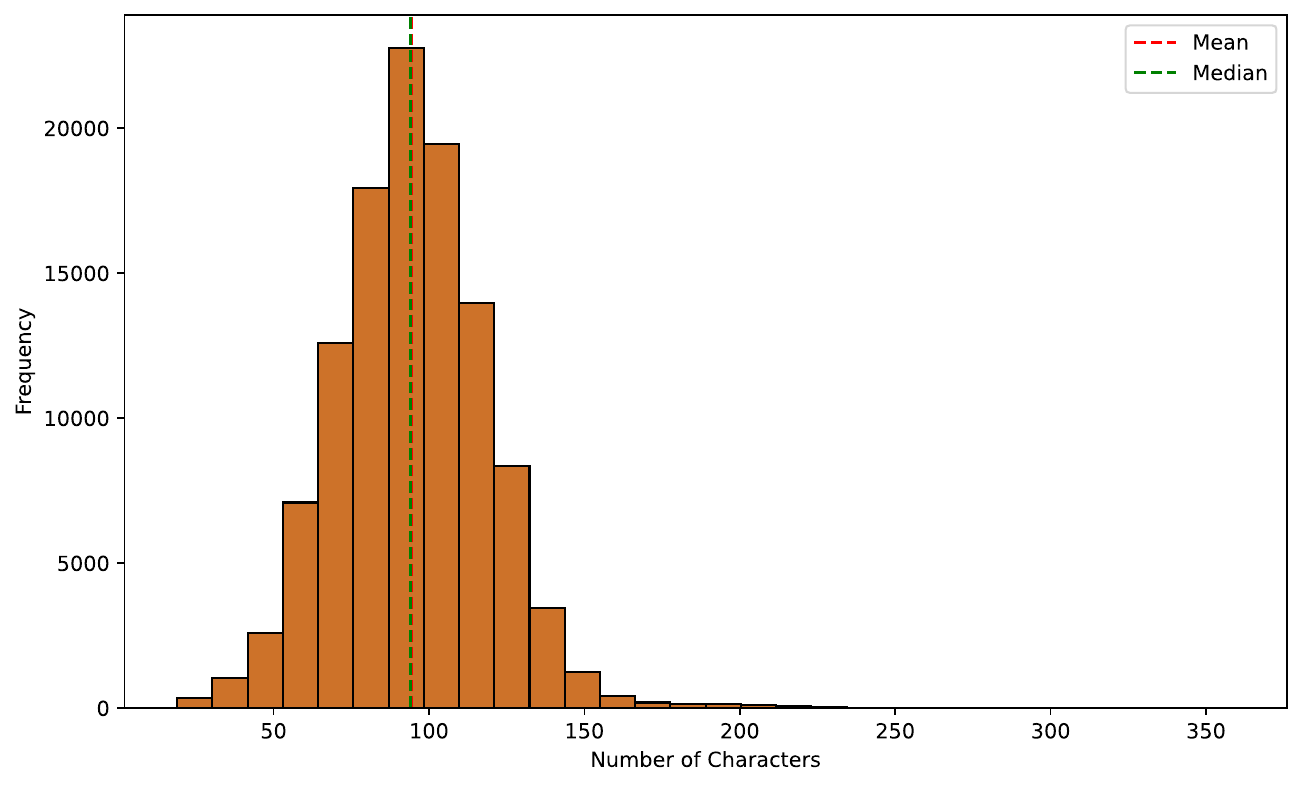}
  \caption{Distribution of headline characters length in dataset}
  \label{fig:HL}
\end{figure}

Graphical analysis of headline length distribution of corpus in our case revealed a maximum of approximately 125 characters as illustrated in Figure~\ref{fig:HL}. 
Therefore, to stay on safe side, we chose a slightly higher number, $\theta = 150$ as the routing threshold. Queries with $\ell(q) < \theta$ are processed through the \textit{headline pathway}, where the query embedding is compared against a pre-computed database of headline embeddings (that also stored full article and category as metadata) using cosine similarity. The top-$k$ most similar headlines are retrieved, and their corresponding full articles, stored as metadata, are recommended to the user. Queries with $\ell(q) \geq \theta$ are routed through the \textit{content pathway},where they are matched against embeddings of full article content (with headline and category as metadata in this case). The top-$k$ articles with the highest similarity scores are then retrieved and recommended to the user.

It is important to clarify that in the headline pathway, the system uses headline embeddings only for similarity matching. The actual recommendations presented to the user are the corresponding full articles. This design exploits the fact that headlines serve as succinct summaries of their entire articles, enabling efficient retrieval of relevant content without requiring to compare against lengthy article bodies. We store a headline embedding along with metadata i.e., full article text and category of each article. Once a headline is a match, the whole article can be recommended immediately.

When an article's content exceeds 512 tokens, we implement a chunking strategy with 50-token overlap. We then take the average of the embeddings of these chunks to generate final embedding of the article. This is the way the entire article is captured.

\subsection{Dimensionality Reduction for Scalable Retrieval}
\label{subsec:dim_reduction}

At a corpus size of approximately 112,000 articles, it was sluggish to search a 768-dimensional space. We considered four dimensionality reduction methods to enhance scalability alongside preserving the semantic relationships: Principal Component Analysis (PCA), Uniform Manifold Approximation and Projection (UMAP), Linear Discriminant Analysis (LDA), and Autoencoders.

\subsubsection{Evaluation Methodology}
We evaluated the performance of each method by its similarity to the original space. For each query, we extracted the top-k results from the reduced dimensional database and compared them to the top-k results of the original 768-dimensional database. The overlap score, which is the number of common articles found in both lists, was used to measure how well the reduced version retained the original retrieval rankings.

PCA, a linear approach that preserves the maximum variance, was more effective than the non-linear alternatives across content and headline embeddings. It is deterministic, fast and provided the maximum scores in overlap. Hence we selected it in both of the retrieval paths. 

\subsubsection{Dimensionality Optimization}
After selecting best DR technique, we adjusted the reduced dimension $d'$ to trade retrieval accuracy with computational efficiency. In the case of content embeddings (long query pathway), we tried 64, 128, and 256 dimensions with 150 to 350 character length queries. The 128 dimensional setting provided the most favorable tradeoff, it retained a large overlap with the original results.

In the case of headline embeddings (short query pathway), we evaluated 98 queries of 100 characters and discovered that 64 was optimal dimension size that maintained the meaning of short headlines without an apparent drop in retrieval quality. The results of dimension-specific optimization indicated that information density differs between lengthy articles and succinct headlines.

Finally, we stored the pooled and reduced embeddings of optimal dimensions into two separate Chroma vector DBs: one containing content column embeddings, and another containing headline column embeddings. 

\subsection{System Architecture and Implementation}
\label{subsec:system_architecture}

The ULTRA framework applies the methodological elements in an integrated pipeline illustrated in Figure \ref{fig:Art}. The system handles user queries through the following steps:

\subsubsection{Query Processing and Routing}

When a user query is entered into the system as $q$, the system first calculates its character length, which is denoted as $\ell(q)$. Depending on the threshold $\theta = 150$  (which again is relative to our recommender system using news corpus and may vary across different recommender systems), the query is routed to the correct pathway. Both paths use the same base transformer model but differ in text source, pooling strategy, and reduced dimension.

\subsubsection{Embedding Generation and Retrieval}
The system uses two distinct vector databases, each specific to the query type. The pipeline applied to a query is dependent on its character length, denoted by $\ell(q)$, relative to the threshold $\theta = 150$, which predetermines the pooling strategy, dimensionality reduction setup, and desired database for similarity search. 

In case queries have length of at least theta, i.e., $\ell(q) \geq \theta$ (long queries), the system will use the content pathway. Mean pooling is used to encode the query text into a 768-dimensional embedding $\mathbf{v}_q$, and then the principal component analysis (PCA) is used to reduce the pooled embeddings to 128 dimensions. This reduced vector $\mathbf{v}'_q$ is compared against a pre-computed database of content embeddings, formed by the same mean pooling and PCA reduction procedure, using cosine similarity. 

In queries whose length is less than the threshold, i.e.,  $\ell(q) < \theta$ (short queries), the system uses the headline pathway. The query is CLS pooled to create its initial embedding which is further shrunk to 64 dimensions using PCA. This succinct representation is compared to a separate set of headline embeddings, all computed identically (CLS pooling + PCA to 64D), to identify the most similar headlines in terms of semantics. 

Both vector databases are stored in ChromaDB with Hierarchical Navigable Small World (HNSW) indexing cofigured for cosine similarity, which allows efficient approximate nearest-neighbor search at scale.

\subsubsection{Result Composition and Presentation}
The  top-$k$  highest ranking articles by cosine similarity are returned together with their metadata (headline, category, and in the case of headline matches, the entire news article). The user is given the final recommendations in a list of relevant news articles with excerpts ranked.

\section{Results and Discussion}
\label{sec_res-discussion}

This section presents a systematic experimental evaluation of our proposed Urdu news recommendation system. We begin in Section~\ref{sub_sec_pooling_eval} by comparing mean, max, and CLS pooling strategies to identify the optimal method for generating article embeddings, revealing distinct performance patterns for short queries (Section~\ref{subsub_sec_short_results}) versus long queries (Section~\ref{subsub_sec_long_results}). Based on the observed under-performance on short queries, we introduce and evaluate headline embeddings as a specialized solution in Section~\ref{sub_sec_headline_analysis}, demonstrating their superiority (Section~\ref{subsub_sec_headline_results}) and analyzing breakthrough cases (Section~\ref{subsub_sec_case_analysis}) that validate our core hypothesis. To address scalability with 112k articles, Section~\ref{sub_sec_dimensionality_reduction} evaluates PCA, UMAP, and Autoencoder for dimensionality reduction, separately analyzing performance on content embeddings (Section~\ref{sub_sec_dr_content}) and headline embeddings (Section~\ref{sub_sec_dr_headline}). Section~\ref{sec_optimal_dimension} determines the optimal reduced dimensions for each pathway i.e., for content embeddings (Section~\ref{sub_sec_optimal_long}) and for headline embeddings (Section~\ref{sub_sec_optimal_short}). Finally, Section~\ref{sub_sec_final_config} presents the complete dual-pathway configuration and evaluates its performance on 100 test queries per pathway (Section~\ref{subsub_sec_final_performance}), followed by comparative analysis with a baseline system (Section~\ref{sec:comparative_analysis}).

\subsection{Evaluation of Pooling Methods for Article Embeddings}
\label{sub_sec_pooling_eval}

To generate fixed-dimensional vector representations from the token-level embeddings produced by the model urduhack/roberta-urdu-small  \cite{51}., we implemented and compared three standard pooling strategies: mean pooling, max pooling, and CLS pooling. Each method aggregates information differently: \textit{mean pooling} computes the average of all token embeddings in the sequence, capturing overall semantic content; \textit{max pooling} selects the maximum value for each dimension across tokens, potentially emphasizing salient features; and CLS pooling utilizes the embedding of the special classification token ([CLS]), which is trained during pre-training to encapsulate sentence-level semantics. 

Before proceeding to results section, it is highly recommended to clarify the distinction between short and long queries definition in our work in the light of use cases, as we have elaborated in Section~\ref{sec_novelty}. 

We used the pre-trained urduhack/roberta-urdu-small transformer model, which generates 768-dimensional token embeddings  \cite{51}.. A fixed maximum sequence length of 512 tokens was employed. For long documents exceeding this limit, a chunking strategy with a 50-token overlap was applied, and the final embedding was computed as the average of all chunk embeddings. Text was tokenized with the model's native tokenizer, applying standard padding and attention masking.

For each pooling method, we created separate vector databases containing embeddings of around 112k Urdu news articles, and stored each article's headline and category as its metadata. To evaluate their retrieval effectiveness, we constructed a test set comprising 18 queries. However it is important to note that, in the last section, we presented a comprehensive evaluation of precision metrics across 100 test queries, thereby demonstrating the efficacy of our proposed methodology. In the current analysis, a smaller subset of queries was employed to delineate the performance disparities among various pooling methods, to facilitate the selection of the most effective one for subsequent implementation.

\begin{figure}[h]
  \centering
  \includegraphics[width=0.9\linewidth, trim=50 460 90 80, clip]{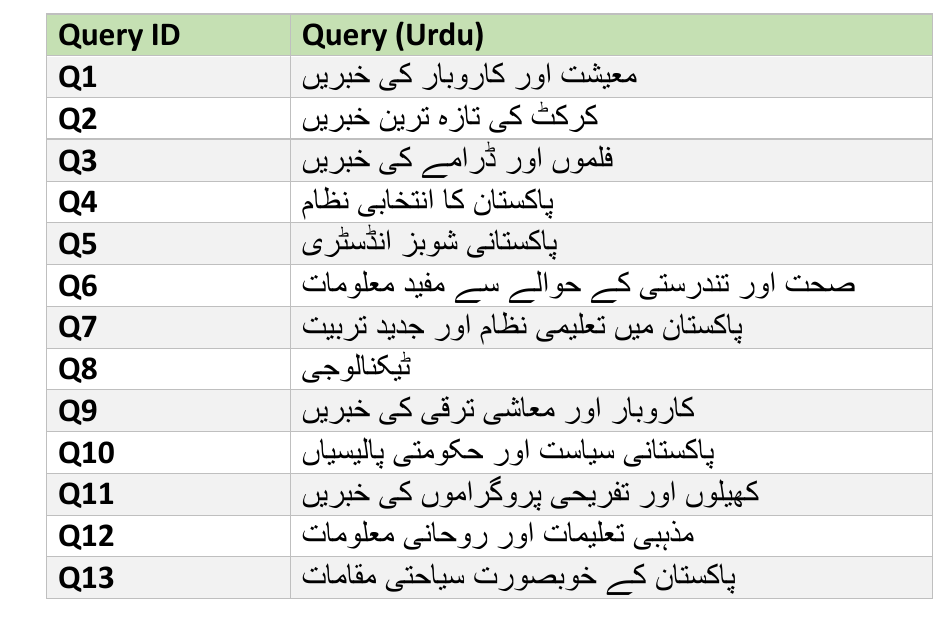}
  \caption{Short length queries}
  \label{fig:short_q}
\end{figure}

\begin{figure}[h]
  \centering
  \includegraphics[width=1\linewidth, trim=50 530 70 70, clip]{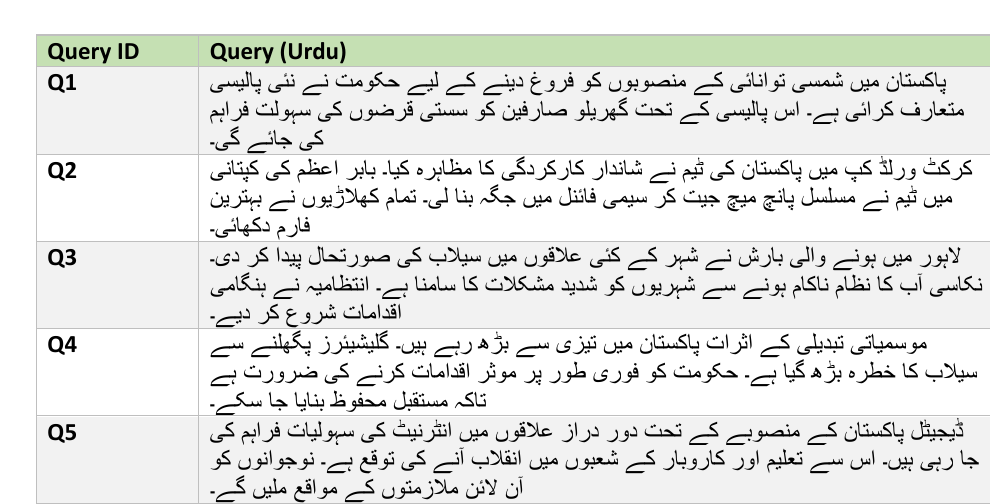}
  \caption{Long length queries}
  \label{fig:long_q}
\end{figure}

Queries were categorized into two distinct types based on length and semantic richness: 13 short queries (user typed search bar queries, concise, keyword-like phrases) as in Figure~\ref{fig:short_q}
and 5 long queries (could be user typed queries or complete articles, detailed, context-rich sentences or paragraphs) as demonstrated in Figure~\ref{fig:long_q}. 

\subsubsection{Performance on Long Queries}
\label{subsub_sec_long_results}

For long, detailed queries, as in Figure~\ref{fig:long_q}, a different performance pattern emerged, as shown in Table~\ref{tab:long_query_results}. Mean pooling achieved the highest average precision at 74\%, followed by max pooling (50\%) and CLS pooling (48\%). 

\begin{table}[htbp]
\centering
\caption{Precision@10 for long queries using different pooling methods on content embeddings}
\label{tab:long_query_results}
\begin{tabular}{>{\raggedright\arraybackslash}p{1cm}cccccc}
\toprule
Query & \multicolumn{2}{c}{Mean Pooling} & \multicolumn{2}{c}{Max Pooling} & \multicolumn{2}{c}{CLS Pooling} \\
\cmidrule(r){2-3} \cmidrule(r){4-5} \cmidrule(r){6-7}
 & TP & Precision & TP & Precision & TP & Precision \\
\midrule
Q1 & 9 & 0.9 (90\%) & 5 & 0.5 (50\%) & 4 & 0.4 (40\%) \\
Q2 & 10 & 1.0 (100\%) & 10 & 1.0 (100\%) & 10 & 1.0 (100\%) \\
Q3 & 4 & 0.4 (40\%) & 1 & 0.1 (10\%) & 0 & 0.0 (0\%) \\
Q4 & 6 & 0.6 (60\%) & 5 & 0.5 (50\%) & 4 & 0.4 (40\%) \\
Q5 & 8 & 0.8 (80\%) & 4 & 0.4 (40\%) & 6 & 0.6 (60\%) \\
\midrule
\textbf{Average} & \textbf{--} & \textbf{0.74}& \textbf{--} & \textbf{0.50}& \textbf{--} & \textbf{0.48 }\\
\textbf{Mean Precision} & & \textbf{74\%}& & \textbf{50\%}& & \textbf{48\%}\\
\bottomrule
\end{tabular}
\end{table}

The superiority of mean pooling for long queries can be attributed to its ability to aggregate semantic information from all tokens in the document. Long queries, which often contain multiple contextual clues and descriptive phrases, benefit from matching against a similarly comprehensive representation of article content. This distributed semantic matching proved more effective than the focused representations from CLS pooling or the potentially noisy salient features from max pooling. Interestingly, in later evaluations with an expanded test set of 100 queries and retrieval depth of $k=15$, mean pooling on content embeddings for long queries achieved precision scores exceeding 90\%, confirming its robustness for detailed semantic matching. 

\subsubsection{Performance on Short Queries}
\label{subsub_sec_short_results}

The Precision@10 results for short queries displayed in Figure~\ref{fig:short_q} is provided in Table~\ref{tab:short_query_results} across the three pooling methods. CLS pooling exhibited better performance with an average precision of 60\%, outperforming mean pooling (53.07\%) and max pooling (28.46\%).

\begin{table*}[htbp]
\centering
\caption{Performance of content embeddings using different pooling methods on short queries (Precision@10)}
\label{tab:short_query_results}
\begin{tabular}{lcccccc}
\toprule
Query & \multicolumn{2}{c}{Mean Pooling} & \multicolumn{2}{c}{Max Pooling} & \multicolumn{2}{c}{CLS Pooling} \\
\cmidrule(r){2-3} \cmidrule(r){4-5} \cmidrule(r){6-7}
 & TP & Precision & TP & Precision & TP & Precision \\
\midrule
Q1& 8 & 0.8 & 1 & 0.1 & 10 & 1.0 \\
Q2& 0 & 0.0 & 1 & 0.1 & 3 & 0.3 \\
Q3& 10 & 1.0 & 10 & 1.0 & 10 & 1.0 \\
Q4& 3 & 0.3 & 0 & 0.0 & 5 & 0.5 \\
Q5& 1 & 0.1 & 3 & 0.3 & 4 & 0.4 \\
Q6& 7 & 0.7 & 10 & 1.0 & 3 & 0.3 \\
Q7& 8 & 0.8 & 3 & 0.3 & 7 & 0.7 \\
Q8& 1 & 0.1 & 0 & 0.0 & 4 & 0.4 \\
Q9& 9 & 0.9 & 0 & 0.0 & 9 & 0.9 \\
Q10& 6 & 0.6 & 0 & 0.0 & 6 & 0.6 \\
Q11& 10 & 1.0 & 9 & 0.9 & 10 & 1.0 \\
Q12& 4 & 0.4 & 0 & 0.0 & 6 & 0.6 \\
Q13& 2 & 0.2 & 0 & 0.0 & 1 & 0.1 \\
\midrule
\textbf{Average} & \textbf{--} & \textbf{0.531} & \textbf{--} & \textbf{0.285} & \textbf{--} & \textbf{0.600} \\
\textbf{Mean Precision} & & \textbf{53.07\%}& & \textbf{28.46\%}& & \textbf{60.00\%}\\
\bottomrule
\end{tabular}
\end{table*}

The performance variance illustrates an important point: although CLS pooling demonstrated the best overall precision with short queries, its best performance (60\%) was not satisfactory in practice, when it comes to recommendations. This means that there is an inherent weakness in applying content-based embeddings to short queries. The semantic expression of a detailed article of 500 words is too broad to effectively correspond to the narrow intent that is captured in a short query.

This continued under-performance on short queries led to the innovative solution presented in the next Section~\ref{sub_sec_headline_analysis}, where we present headline embeddings that are specifically created to address this semantic gap.

\subsection{Bridging the Semantic Gap: Headline Embeddings for Short Queries}
\label{sub_sec_headline_analysis}

The highest precision of 60\% on short queries, even using the most effective pooling method on content embeddings, was insufficient to a production-scale recommendation system. This gap in performance is caused by a mismatch in semantic granularity: short queries reflect narrow user intent, and content embeddings reflect the multidimensional themes of full-length articles. This observation inspired the design of our dual-embedding strategy, which aims to match query semantics with relevant document representations.

\subsubsection{Hypothesis and Implementation}
\label{subsub_sec_hypothesis}

Our hypothesis was that short user queries were semantically closer to article headlines than to the content of full articles. Headlines are, by definition, short, information-dense summaries of the essence of an article, and are therefore more semantically aligned to short queries. To test this hypothesis, we produced a new set of embeddings of the headline column of our dataset using the same pre-trained model (urduhack/roberta-urdu-small  \cite{51}). For all pooling techniques (mean, max, and CLS), we prepared individual pooling vector databases in which every headline embedding was stored with its respective full article text and category as metadata. This architecture guaranteed that in case a headline corresponded with a brief query in semantic terms, the system would be able to fetch and suggest the entire article to the user, not just headline.

A key difference in processing was that headlines were short enough to not need chunking and could be processed in single forward pass of the transformer model whereas long article text needed chunking outlined earlier.

\subsubsection{Performance Evaluation on Short Queries}
\label{subsub_sec_headline_results}

The evaluation focused exclusively on the same 13 short queries as illustrated in Figure~\ref{fig:short_q} that were used for evaluation of pooling methods for content embeddings, as our intuition posited that the semantic compatibility between short queries and headlines would only be beneficial for this query type. For long queries, content embeddings remained the appropriate choice due to their comprehensive semantic coverage. 

Table~\ref{tab:headline_short_results} presents the Precision@10 results for short queries shown in Figure~\ref{fig:short_q} across the three pooling methods. CLS pooling demonstrated superior performance, significantly outperforming mean and max pooling.

\begin{table}[htbp]
\centering
\caption{Performance of headline embeddings using different pooling methods on short queries (Precision@10)}
\label{tab:headline_short_results}
\resizebox{\columnwidth}{!}{%
\begin{tabular}{@{}lcccccc@{}}
\toprule
\multirow{2}{*}{Query} & \multicolumn{2}{c}{Mean Pooling} & \multicolumn{2}{c}{Max Pooling} & \multicolumn{2}{c}{CLS Pooling} \\
\cmidrule(lr){2-3} \cmidrule(lr){4-5} \cmidrule(l){6-7}
 & TP & Prec. & TP & Prec. & TP & Prec. \\
\midrule
Q1 & 8 & 0.8 & 4 & 0.4 & 9 & 0.9 \\
Q2 & 6 & 0.6 & 3 & 0.3 & 9 & 0.9 \\
Q3 & 7 & 0.7 & 10 & 1.0 & 10 & 1.0 \\
Q4 & 1 & 0.1 & 0 & 0.0 & 8 & 0.8 \\
Q5 & 6 & 0.6 & 6 & 0.6 & 9 & 0.9 \\
Q6 & 7 & 0.7 & 8 & 0.8 & 9 & 0.9 \\
Q7 & 7 & 0.7 & 5 & 0.5 & 10 & 1.0 \\
Q8 & 2 & 0.2 & 0 & 0.0 & 9 & 0.9 \\
Q9 & 9 & 0.9 & 9 & 0.9 & 9 & 0.9 \\
Q10 & 4 & 0.4 & 5 & 0.5 & 8 & 0.8 \\
Q11 & 9 & 0.9 & 6 & 0.6 & 9 & 0.9 \\
Q12 & 3 & 0.3 & 3 & 0.3 & 9 & 0.9 \\
Q13 & 7 & 0.7 & 4 & 0.4 & 6 & 0.6 \\
\textbf{Average}& & \textbf{0.5846}& & \textbf{0.4846}& & \textbf{0.8769}\\
\multicolumn{1}{c}{\textbf{Mean Precision}} & & \textbf{58.46\%}& & \textbf{48.46\%}& & \textbf{87.69\%}\\
\bottomrule
\end{tabular}%
}
\end{table}

\paragraph{Superiority of CLS Pooling on Headline Embeddings}
For headline-based embeddings, CLS pooling achieved a remarkable average precision of 87.69\%, significantly outperforming both mean pooling (58.46\%) and max pooling (48.46\%) on the same text source. This represents a 46.2\% relative improvement over the best content-based method (CLS pooling at 60\%). We posit that because headlines are short and syntactically focused, the [CLS] token, which is specifically trained during pre-training to capture sentence-level semantics, effectively encapsulates the entire headline's meaning without the dilution that can occur from averaging across tokens (mean pooling) or the potential introduction of noise from selecting maximum values (max pooling).

\subsubsection{Case Analysis: Precision Breakthroughs}
\label{subsub_sec_case_analysis}

\begin{figure}[h]
  \centering
  \includegraphics[width=0.9\linewidth, trim=50 440 0 70, clip]{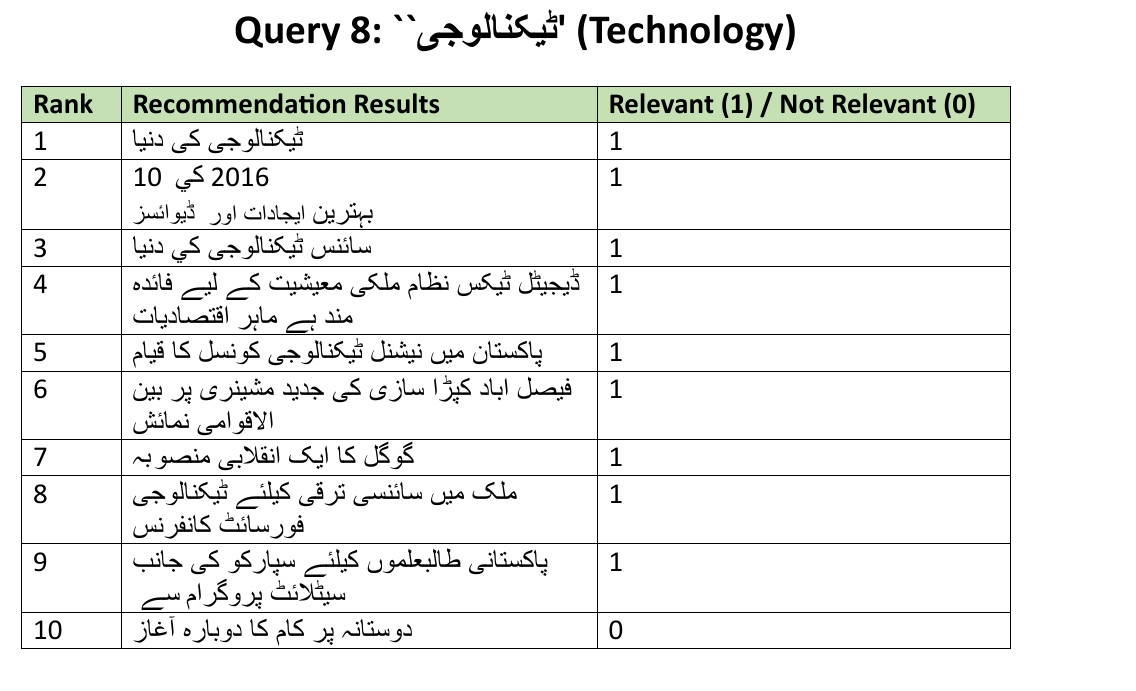}
  \caption{Urdu short query 8 recommendations example 1}
  \label{fig:q_8_r}
\end{figure}

The headline embedding approach, particularly with CLS pooling, enabled breakthroughs on queries where other methods failed catastrophically. A few such examples of precision breakthroughs are discussed below.
The single-word query in Figure~\ref{fig:q_8_r} represents the most challenging case for semantic matching. Content embeddings struggled with this query, achieving only 40\% precision with CLS pooling. In contrast, CLS-pooled headline embeddings achieved 90\% precision corresponding to 9 TPs out of 10, retrieving highly relevant technology-focused headlines such as shown in Figure~\ref{fig:q_8_r}, where 1 indicates relevant query, and 0 indicates irrelevant.

\begin{figure}[h]
  \centering
  \includegraphics[width=0.9\linewidth, trim=50 420 0 70, clip]{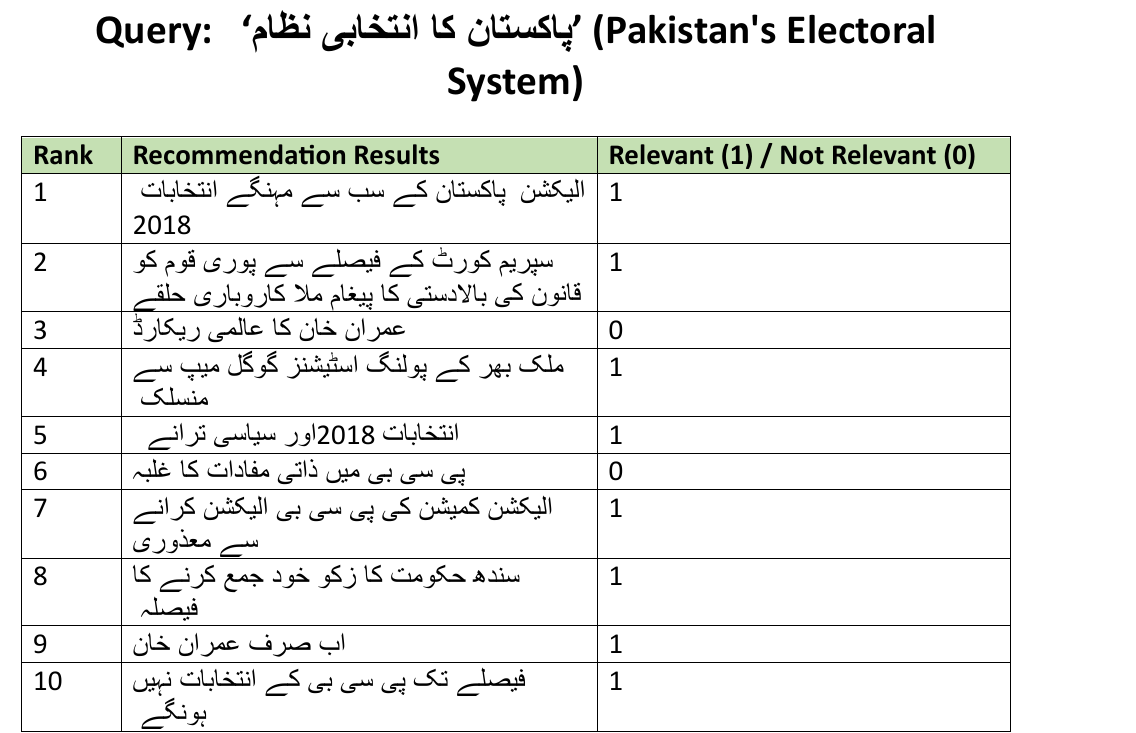}
  \caption{Urdu short query recommendations example 2}
  \label{fig:q_4_r}
\end{figure}

Similarly another short query as in  Figure~\ref{fig:q_4_r} exemplifies how headline embeddings better capture domain-specific semantics. With content embeddings, precision was only 50\%, but headline embeddings with CLS pooling achieved 80\% precision meaning there were 8 TPs, correctly retrieving election-related headlines as shown in Figure~\ref{fig:q_4_r}, where 1 indicates relevant query, and 0 indicates irrelevant.

\begin{figure}[h]
  \centering
  \includegraphics[width=0.9\linewidth, trim=20 460 0 70, clip]{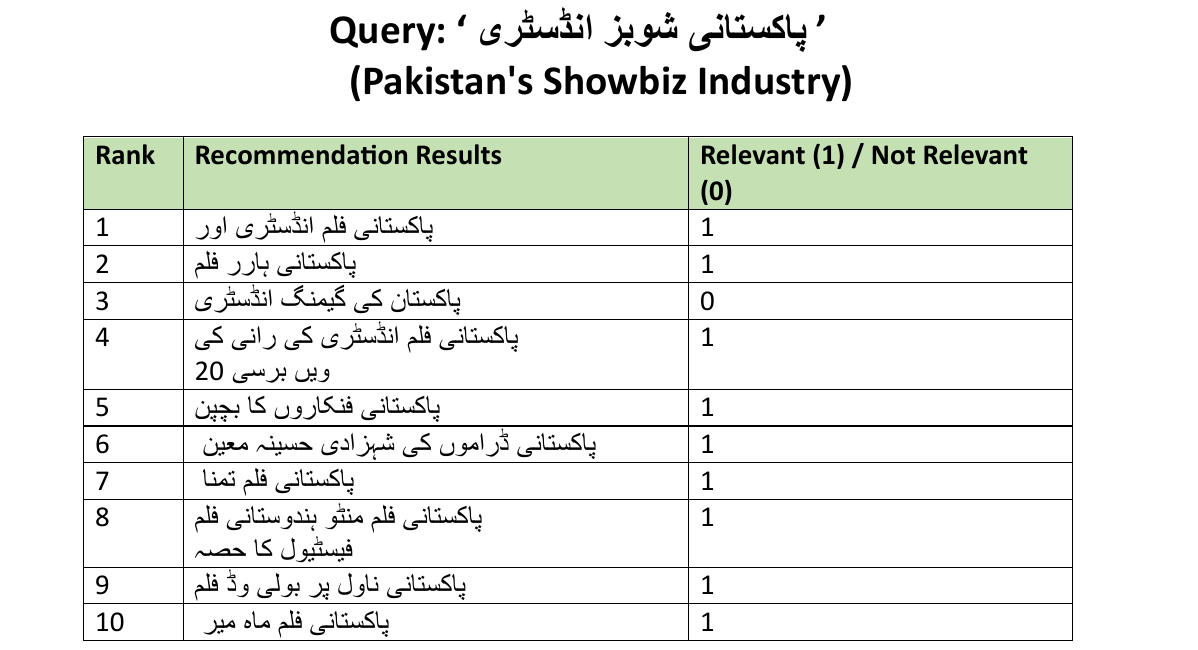}
  \caption{Urdu short query recommendations example 3}
  \label{fig:q_5_r}
\end{figure}

Another query as shown in Figure~\ref{fig:q_5_r} demonstrates the improved specificity of headline embeddings. Previous methods retrieved general entertainment news from Bollywood and Hollywood due to the broad semantic representation in content embeddings. However, CLS-pooled headline embeddings correctly captured the nationality modifier "Pakistani", retrieving exclusively Pakistan-centric entertainment headlines achieving 90\% precision as illustrated in Figure~\ref{fig:q_5_r}, where 1 indicates relevant query, and 0 indicates irrelevant.

Table~\ref{tab:short_results} provides a comprehensive analysis illustrating the dramatic improvements in precision for recommendation results generated in response to short queries, when employing CLS pooling on headline embeddings (which outperformed in this scenario) compared to CLS pooling on content embeddings (which excelled in the alternative case). 

\begin{figure}[h]
  \centering
  \includegraphics[width=1\linewidth, trim=10 330 20 65, clip]{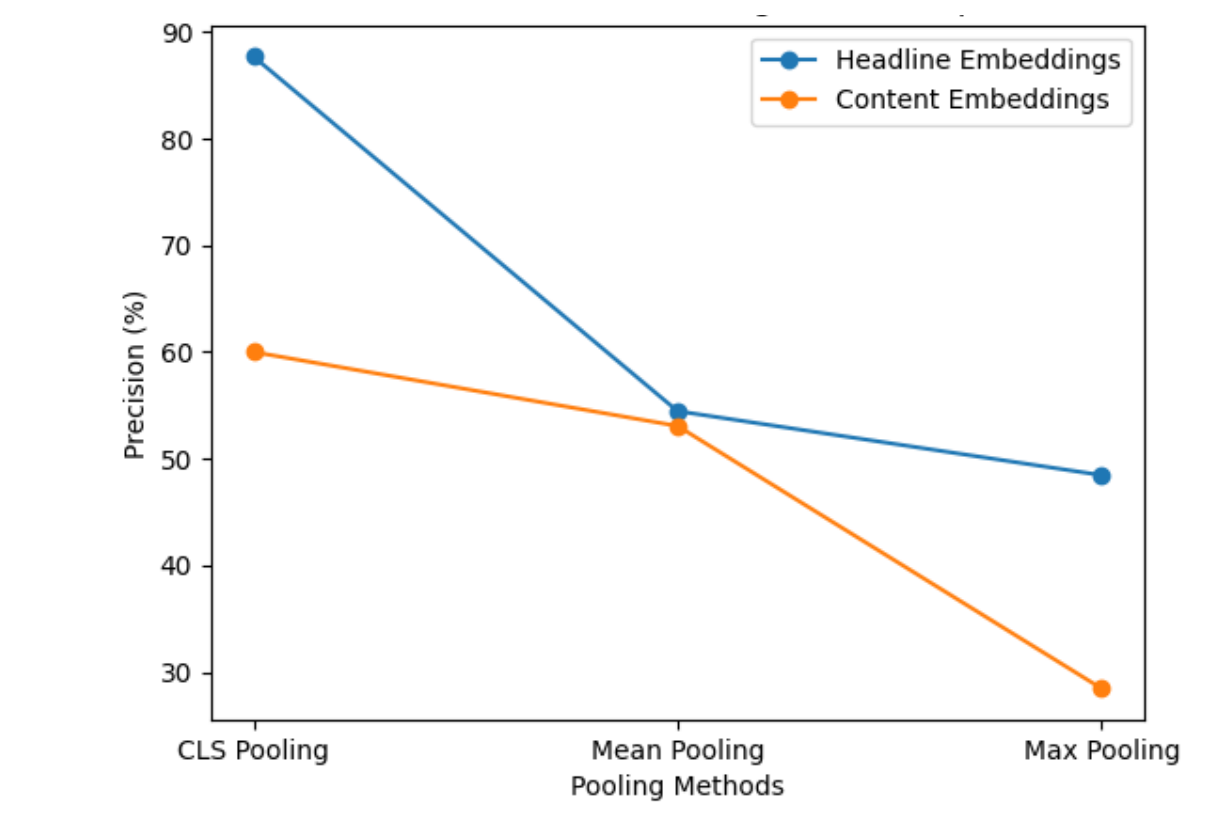}
  \caption{Comparison of precision@10 for short queries of all pooling methods for both headline and content embeddings}
  \label{fig:short_q_v_h_g}
\end{figure}

\begin{table}[htbp]
\centering
\caption{Comparative Precision@10 for short queries: Content vs. headline embeddings using CLS pooling method}
\label{tab:short_results}
\resizebox{\columnwidth}{!}{%
\begin{tabular}{@{}lcccccc@{}}
\toprule
\multirow{2}{*}{Query} & \multicolumn{2}{c}{Content (CLS)} & \multicolumn{2}{c}{Headline (CLS)} & \multirow{2}{*}{Imp.} \\
\cmidrule(lr){2-3} \cmidrule(l){4-5}
 & TP & Prec. & TP & Prec. & \\
\midrule
Q1 & 10 & 1.00 & 9 & 0.90 & $\mathbf{-10.0\%}$ \\
Q2 & 3 & 0.30 & 9 & 0.90 & $\mathbf{+200.0\%}$ \\
Q3 & 10 & 1.00 & 10 & 1.00 & $\mathbf{0.0\%}$ \\
Q4 & 5 & 0.50 & 8 & 0.80 & $\mathbf{+60.0\%}$ \\
Q5 & 4 & 0.40 & 9 & 0.90 & $\mathbf{+125.0\%}$ \\
Q6 & 3 & 0.30 & 9 & 0.90 & $\mathbf{+200.0\%}$ \\
Q7 & 7 & 0.70 & 10 & 1.00 & $\mathbf{+42.9\%}$ \\
Q8 & 4 & 0.40 & 9 & 0.90 & $\mathbf{+125.0\%}$ \\
Q9 & 9 & 0.90 & 9 & 0.90 & $\mathbf{0.0\%}$ \\
Q10 & 6 & 0.60 & 8 & 0.80 & $\mathbf{+33.3\%}$ \\
Q11 & 10 & 1.00 & 9 & 0.90 & $\mathbf{-10.0\%}$ \\
Q12 & 6 & 0.60 & 9 & 0.90 & $\mathbf{+50.0\%}$ \\
Q13 & 1 & 0.10 & 6 & 0.60 & $\mathbf{+500.0\%}$ \\
\midrule
\textbf{Avg. Prec.} & \multicolumn{2}{c}{\textbf{0.600}} & \multicolumn{2}{c}{\textbf{0.877}} & \\
\textbf{Overall Imp.} & \multicolumn{2}{c}{\textbf{--}} & \multicolumn{2}{c}{\textbf{--}} & \textbf{+46.2\%} \\
\bottomrule
\end{tabular}%
}
\end{table}
Figure~\ref{fig:short_q_v_h_g} shows the line chart comparison of all pooling methods using both content column and headline column embeddings.
\subsubsection{Implications for Pooling Strategy Selection}
\label{subsub_sec_pooling_implications}

The experimental results yield clear guidelines for pooling strategy selection in our dual-embedding framework:

\begin{equation}
\text{Pooling Strategy} = 
\begin{cases}
\text{Mean Pooling for Content Embeddings}, \\
\text{CLS Pooling for Headline Embeddings}
\end{cases}
\end{equation}

This is an optimized design that exploits the advantages of both designs: comprehensive aggregation of mean pooling to match long and detailed queries to rich article content achieving the best results (74\% precision) and focused semantic capture of CLS pooling to match short and purposeful queries to brief headlines. The 87.69\% precision on short queries is not only a quantitative change, but a qualitative shift in how the system comprehends and acts on focused user intent. 

The success of this approach\textbf{ validates our core hypothesis}: aligning the semantic granularity of the query with an appropriately granular document representation is paramount for recommendation relevance. This insight forms the foundation of our proposed \textbf{dual-embedding architecture}.

\subsection{Dimensionality Reduction for Efficient Retrieval}
\label{sub_sec_dimensionality_reduction}

Although our dual-embedding methodology created optimal structure for semantic matching, there was still a concern on operational efficiency. With extremely large corpus size, nearest-neighbor searches in the original 768-dimensional space were very expensive in terms of computational and time costs. To improve the scalability of our recommendation system and its real-time responsiveness without significantly reducing the quality of retrieval, we explored the methods of dimensionality reduction (DR). It attempted to identify lower-dimensional representations that could maintain the key semantic associations within both content and headline embedding spaces.

\subsubsection{Dimensionality reduction for content embeddings}
\label{sub_sec_dr_content}

Given our established dual-embedding architecture as discussed in Section~\ref{subsub_sec_pooling_implications} , we first focused on the content embedding space, where mean-pooled 768-dimensional vectors represented full news articles. To identify the most suitable DR technique, we implemented and compared three prominent methods: Principal Component Analysis (PCA), Uniform Manifold Approximation and Projection (UMAP), and a neural Autoencoder.

\paragraph{Methodology and configurations}
\label{dr_methodology}

All techniques were configured to reduce embeddings from 768 dimensions to 64 dimensions, enabling a fair and direct comparison of their ability to preserve semantic information. The configurations were as follows:
\begin{itemize}
    \item \textbf{PCA (Principal Component Analysis):} A linear, unsupervised technique configured with \texttt{n\_components=64} to capture the directions of maximal variance in the original data.
    \item \textbf{UMAP (Uniform Manifold Approximation and Projection):} A non-linear manifold learning method with hyperparameters \texttt{n\_neighbors=15}, \texttt{min\_dist=0.1}, and \texttt{metric='cosine'} to model the underlying topological structure.
    \item \textbf{Autoencoder:} A symmetric neural network with encoder-decoder architecture (768 → 384 → 192 → 64 → 192 → 384 → 768). It was trained for 50 epochs using Mean Squared Error loss, the Adam optimizer (learning rate=0.001), employed Dropout (0.2) and Batch Normalization for regularization.
\end{itemize}

The selection of $K=50$ for evaluation was deliberate; while PCA and the Autoencoder showed comparable performance at smaller $K$ values (e.g., $K=10$), increasing the retrieval depth helped us expose which DR technique has superior ability to maintain ranking fidelity across a broader spectrum of similar items. The dimensionality reduction was applied to the 768-dimensional mean-pooled content embeddings, as mean pooling was established in Section ~\ref{subsub_sec_pooling_implications} as the optimal strategy for generating semantic representations of full-length articles.

\paragraph{Evaluation strategy}
\label{subsub_sec_dr_evaluation}

The performance of each DR method was evaluated by its fidelity to the original high-dimensional semantic space. For a set of 13 diverse, long-form test queries (representative of the "long query" pathway), we retrieved the top-50 recommendations ($K=50$) from both the original 768-dimensional database (serving as ground truth) and each 64-dimensional reduced database. 
The key metric was the Overlap Score—the number of articles common to both result sets—expressed both as a raw count and a percentage of the retrieved items.  We also computed the Jaccard Index ($J = |A \cap B| / |A \cup B|$) to measure the similarity between the recommendation sets. For a query $q$, with $A_q$ as the top-$k$ results from the original 768D space and $B_q$ as the top-$k$ results from the reduced dimensional space, the Jaccard Index at retrieval depth $k$ is defined as:

\begin{equation}
\label{eq:jaccard_at_k}
J@k(q) = \frac{|A_q \cap B_q|}{|A_q \cup B_q|} = \frac{\text{Overlap}_q}{2k - \text{Overlap}_q}
\end{equation}

The overall effectiveness of a dimensionality reduction technique was then quantified by averaging this index across all $Q$ test queries, yielding the Mean Jaccard Index at $k$:

\begin{equation}
\label{eq:mean_jaccard_at_k}
\text{Mean-}J@k = \frac{1}{Q} \sum_{q=1}^{Q} J@k(q)
\end{equation}

This measure gives a normalized quantification of set similarity taking into consideration true positives and the union of retrieved items, which is a strong measure of the extent to which the reduced space maintains the retrieval properties of the original embedding space.

\paragraph{Performance of DR techniques on content embeddings}
\label{dr_results_content}

\begin{figure}[h]
  \centering
  \includegraphics[width=1\linewidth, trim=0 0 0 0, clip]{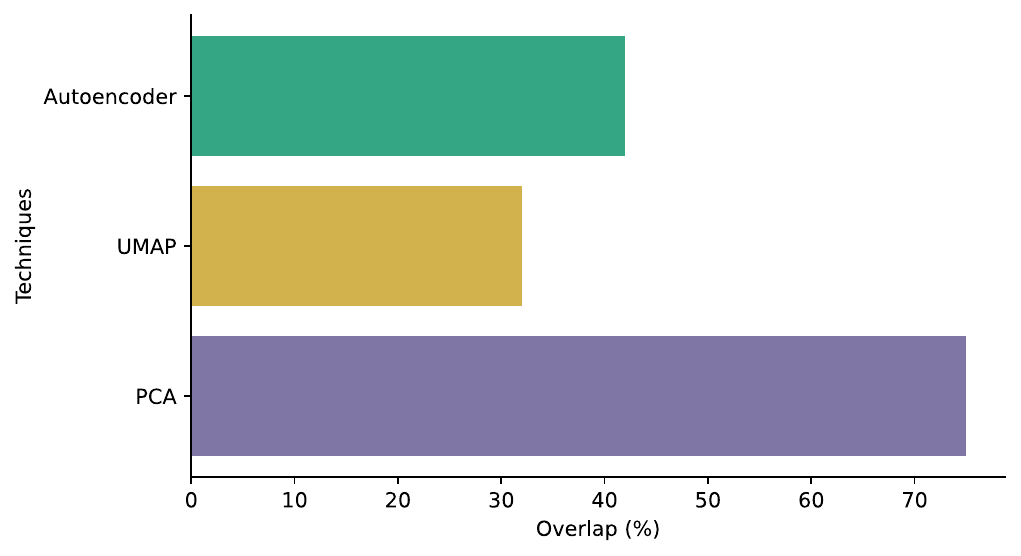}
  \caption{Fidelity of 64-dimensional embeddings to the 768-dimensional ground truth: average overlap in top-50 recommendation across DR techniques for content vector DB}
  \label{fig:overlap}
\end{figure}

\begin{figure}[h]
  \centering
  \includegraphics[width=0.7\linewidth, trim=0 0 0 0, clip]{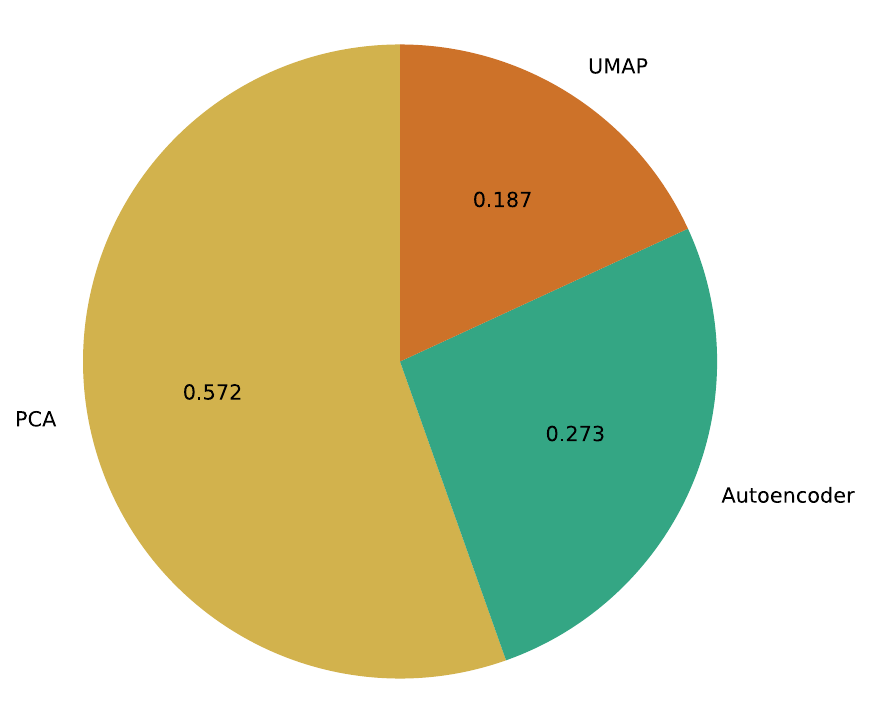}
  \caption{Comparison of average Jaccard Index distributions to analyze fidelity of DR technique to original semantic space for content vector DB}
  \label{fig:jaccard}
\end{figure}

Table~\ref{tab:dr_content_results} presents the comparative performance of the three DR techniques on content embeddings. PCA demonstrated decisive superiority, achieving a mean overlap score of 74.92\% and a mean Jaccard Index of 0.572 as shown in Figures~\ref{fig:overlap} and~\ref{fig:jaccard}.

\begin{table}[ph!]
\FloatBarrier
\caption{Performance of DR techniques on content embeddings (Top-50 Retrieval)}
\label{tab:dr_content_results}
\resizebox{\columnwidth}{!}{%
\begin{tabular}{@{}lccccccccc@{}}
\toprule
\multirow{2}{*}{Query} & \multicolumn{3}{c}{Overlap with 768D Ground Truth} & \multicolumn{3}{c}{Overlap Count} & \multicolumn{3}{c}{Jaccard Index} \\
\cmidrule(lr){2-4} \cmidrule(lr){5-7} \cmidrule(l){8-10}
 & PCA & UMAP & AE & PCA & UMAP & AE & PCA & UMAP & AE \\
\midrule
Q1 & 86.0\% & 46.0\% & 64.0\% & 43 & 23 & 32 & 0.754 & 0.299 & 0.471 \\
Q2 & 86.0\% & 22.0\% & 36.0\% & 43 & 11 & 18 & 0.754 & 0.124 & 0.220 \\
Q3 & 90.0\% & 46.0\% & 52.0\% & 45 & 23 & 26 & 0.818 & 0.299 & 0.351 \\
Q4 & 62.0\% & 10.0\% & 16.0\% & 31 & 5 & 8 & 0.449 & 0.053 & 0.087 \\
Q5 & 70.0\% & 46.0\% & 0.0\% & 35 & 23 & 0 & 0.538 & 0.299 & 0.000 \\
Q6 & 42.0\% & 16.0\% & 18.0\% & 21 & 8 & 9 & 0.266 & 0.087 & 0.099 \\
Q7 & 82.0\% & 32.0\% & 44.0\% & 41 & 16 & 22 & 0.695 & 0.190 & 0.282 \\
Q8 & 78.0\% & 26.0\% & 44.0\% & 39 & 13 & 22 & 0.639 & 0.149 & 0.282 \\
Q9 & 66.0\% & 28.0\% & 48.0\% & 33 & 14 & 24 & 0.493 & 0.163 & 0.316 \\
Q10 & 92.0\% & 46.0\% & 78.0\% & 46 & 23 & 39 & 0.852 & 0.299 & 0.639 \\
Q11 & 66.0\% & 16.0\% & 26.0\% & 33 & 8 & 13 & 0.493 & 0.087 & 0.149 \\
Q12 & 76.0\% & 50.0\% & 74.0\% & 38 & 25 & 37 & 0.613 & 0.333 & 0.587 \\
Q13 & 78.0\% & 38.0\% & 50.0\% & 39 & 19 & 25 & 0.639 & 0.235 & 0.333 \\
\bottomrule
\end{tabular}%
}
\end{table}

Figures~\ref{fig:overlap} and~\ref{fig:jaccard} present the evaluation results of dimensionality reduction techniques applied to content embeddings evaluated on long test queries. Figure~\ref{fig:overlap} shows a horizontal bar graph of the average of top-50 recommendation overlap between 64-dimensional reduced embeddings and the 768-dimensional ground truth, while Figure~\ref{fig:jaccard} displays the corresponding average Jaccard index distribution across techniques.

The results reveal a clear hierarchy: PCA significantly outperformed both UMAP and the Autoencoder. The PCA linear projection was extremely successful in maintaining the semantic neighborhood structure that is important in ranking a large number of similar articles. Although the Autoencoder was performing reasonably on multiple queries, it was quite inconsistent (e.g. Q5 with 0\% overlap). UMAP, although it is strong at visualizing complex manifolds, had the worst performance in this retrieval-based task, suggesting that its non-linear transformations can lead to distortion of the global similarity relationships critical to a successful nearest-neighbor search.

\subsubsection{Dimensionality reduction for headline embeddings}
\label{sub_sec_dr_headline}

After the final findings on the content embeddings, we applied the same evaluation systematically to headline embeddings space which is the crucial component of our short query pathway. The same three methods (PCA, UMAP, and Autoencoder) were used to further reduce the 768-dimensional CLS-pooled headline embeddings to 64 dimensions since CLS pooling was determined in Section ~\ref{subsub_sec_pooling_implications} as the best approach to provide semantic representations of headlines. The remaining configurations are same as were adopted in case of DR analysis for content embeddings, discussed in Section ~\ref{dr_methodology}. The evaluation utilized the same 13 short queries as shown in Figure~\ref{fig:short_q} that defined the short query category, with the original 768D headline embeddings serving as the ground truth. Retrieval depth was set to $K=50$ to provide a rigorous test of each technique's ability to preserve ranking fidelity across a broad set of similar items.

\subsubsection{Results and Comparative Analysis}
\label{subsub_sec_headline_dr_results}

\begin{figure}[h]
  \centering
  \includegraphics[width=0.9\linewidth, trim=0 0 0 0, clip]{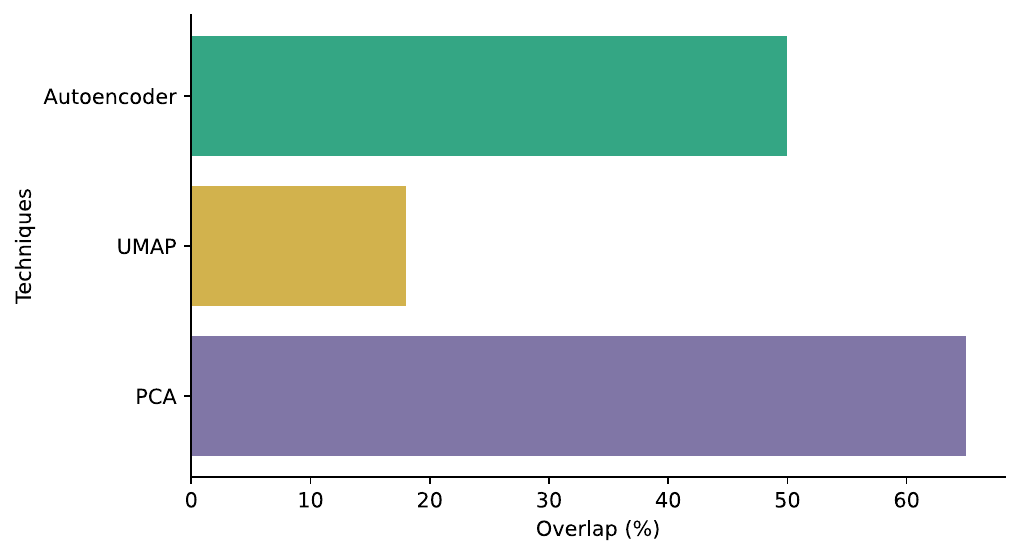}
  \caption{Fidelity of 64-dimensional embeddings to the 768-dimensional ground truth for headline vector DB: average overlap in top-50 across DR techniques}
  \label{fig:overlap2}
\end{figure}

\begin{figure}[h]
  \centering
  \includegraphics[width=0.7\linewidth, trim=0 0 0 0, clip]{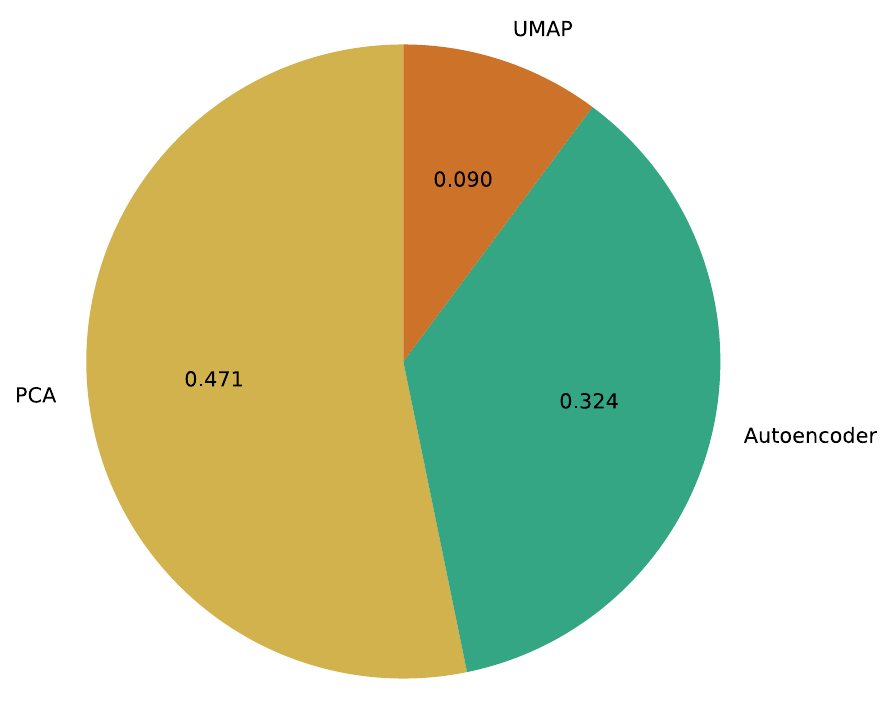}
  \caption{Comparison of average Jaccard Index distributions to analyze fidelity of DR technique to original semantic space for headline vector DB}
  \label{fig:jaccard2}
\end{figure}

The performance metrics for headline embeddings are presented in Table~\ref{tab:dr_headline_results}. PCA once again demonstrated clear superiority, achieving a mean overlap score of 65.10\% and a mean Jaccard Index of 0.471 illustrated in Figures~\ref{fig:overlap2} and~\ref{fig:jaccard2}.

\begin{table}[htbp]
\centering
\caption{Performance of DR techniques on headline embeddings (Top-50 Retrieval)}
\label{tab:dr_headline_results}
\resizebox{\columnwidth}{!}{%
\begin{tabular}{@{}lccccccccc@{}}
\toprule
\multirow{2}{*}{Query} & \multicolumn{3}{c}{Overlap with 768D Ground Truth} & \multicolumn{3}{c}{Overlap Count} & \multicolumn{3}{c}{Jaccard Index} \\
\cmidrule(lr){2-4} \cmidrule(lr){5-7} \cmidrule(l){8-10}
 & PCA & UMAP & AE & PCA & UMAP & AE & PCA & UMAP & AE \\
\midrule
Q1 & 46.0\% & 10.0\% & 32.0\% & 23 & 5 & 16 & 0.299 & 0.053 & 0.190 \\
Q2 & 20.0\% & 14.0\% & 18.0\% & 10 & 7 & 9 & 0.111 & 0.075 & 0.099 \\
Q3 & 80.0\% & 30.0\% & 68.0\% & 40 & 15 & 34 & 0.667 & 0.176 & 0.515 \\
Q4 & 62.0\% & 26.0\% & 40.0\% & 31 & 13 & 20 & 0.449 & 0.149 & 0.250 \\
Q5 & 76.0\% & 20.0\% & 58.0\% & 38 & 10 & 29 & 0.613 & 0.111 & 0.408 \\
Q6 & 82.0\% & 24.0\% & 50.0\% & 41 & 12 & 25 & 0.695 & 0.136 & 0.333 \\
Q7 & 86.0\% & 12.0\% & 68.0\% & 43 & 6 & 34 & 0.754 & 0.064 & 0.515 \\
Q8 & 66.0\% & 12.0\% & 58.0\% & 33 & 6 & 29 & 0.493 & 0.064 & 0.408 \\
Q9 & 42.0\% & 14.0\% & 36.0\% & 21 & 7 & 18 & 0.266 & 0.075 & 0.220 \\
Q10 & 66.0\% & 14.0\% & 50.0\% & 33 & 7 & 25 & 0.493 & 0.075 & 0.333 \\
Q11 & 74.0\% & 18.0\% & 54.0\% & 37 & 9 & 27 & 0.587 & 0.099 & 0.370 \\
Q12 & 84.0\% & 14.0\% & 62.0\% & 42 & 7 & 31 & 0.724 & 0.075 & 0.449 \\
Q13 & 62.0\% & 20.0\% & 62.0\% & 31 & 10 & 31 & 0.449 & 0.111 & 0.449 \\
\bottomrule 
\end{tabular}%
}
\end{table}
Figures~\ref{fig:overlap2} and~\ref{fig:jaccard2} present the evaluation results of dimensionality reduction techniques applied to headline embeddings evaluated on short test queries defined in Figure~\ref{fig:short_q} . Figure~\ref{fig:overlap2} shows a horizontal bar graph of the average top-50 recommendation overlap between 64-dimensional `reduced embeddings and the 768-dimensional ground truth, while Figure~\ref{fig:jaccard2} displays the corresponding average Jaccard index distribution across techniques.

\subsubsection{Selection of Dimensionality Reduction Technique}
\label{subsub_sec_dr_selection}

The Autoencoder showed improved relative performance on headline embeddings compared to content embeddings, yet remained inconsistent. UMAP continued to perform poorly, confirming its unsuitability for preserving global similarity structures in this retrieval context. PCA's robust performance across two distinct textual domains (long-form content and short headlines) and two different optimal pooling strategies (mean and CLS) demonstrates its generalizability and reliability as a dimensionality reduction technique for semantic embedding spaces in information retrieval tasks.

Based on the comprehensive evaluation across both content and headline embeddings, Principal Component Analysis (PCA) was selected as the optimal dimensionality reduction technique for our dual-embedding architecture. The absolute overlap scores for headline embeddings (mean 65.10\%) and content embeddings (mean 74.92\%). The adoption of PCA enables efficient storage and rapid similarity search in our production system while preserving approximately 65-75\% of the original semantic retrieval fidelity, as measured by overlap with full-dimensional results.

\subsection{Determining Optimal Embedding Dimension}
\label{sec_optimal_dimension}

Following the selection of Principal Component Analysis (PCA) as the superior dimensionality reduction technique, we sought to identify the optimal reduced dimension size for each pathway of our dual-embedding architecture. The primary objective was to maximize the preservation of semantic fidelity relative to the original 768-dimensional space while achieving a practical dimensionality for efficient storage and retrieval. This is an inevitable trade-off: reducing dimensions can result in an increased loss of information, whereas increasing dimensions is a diminishing game in terms of semantic preservation, and raising the cost of computation and storage. 

To comprehensively investigate this trade-off, we considered three candidate dimensions, 64, 128, and 256 chosen as powers of 2 to match to computational best practices in the area of vector processing. The analysis was done on both content embeddings (serving long queries) and headline embeddings (serving short queries) to determine dimension-specialized optima in each semantic space. 

\subsubsection{Optimal Dimension for Long Queries: Content Embeddings}
\label{sub_sec_optimal_long}

In the above dimensionality reduction analysis, all candidate techniques (PCA, UMAP, Autoencoder) used a target dimension of 64D to guarantee a fair and balanced comparison for regular assessment of their capacity to maintain semantic information. But to make system practical, it is necessary to identify the particular reduced dimension that optimizes speed of retrieval and semantic fidelity to every embedding pathway. This trade-off is implicit: lower dimensions imply faster computations but often lead to higher loss of information, while higher dimensions retain semantics at higher computation costs. 

In our dual-embedding architecture query routing is controlled by a character-level cutoff, denoted by $\theta$, as established in
Section~\ref{sub_sub_sec_adaptive_routing}. The threshold $\theta$ = 150 characters was determined empirically based on the maximum headline length in our corpus, ensuring that queries semantically comparable to headlines are routed to the headline embedding pathway. This threshold selection principle is generalizable: $\theta$ should be informed by the characteristic length of titles, headings, or product titles in a given recommender system’s content corpus.

It is important to clarify the operational definition of a
"long query" in our system. In addition to conventional short search- bar queries, we regard the entire text of an article a user is reading actively as a long query. Such a situation is typical in content recommendation systems, in which the article currently viewed acts as the contextual seed to produce related recommendations. In this instance, the whole content of the article acts as query and activates the long-query pipeline. Empirical evidence shows that the relevance of recommendations tends to rise with the length of query, with a more informative context permitting it to better match semantically.

The lower bound for this pathway is $\theta$ = 150 characters.
To ensure system robustness across the natural variation
in article lengths, our evaluation for dimension selection
was structured to test performance across a spectrum of
long-query sizes. We constructed a test set of 100 queries at each of five distinct length intervals: 150, 200, 250, 300, and 350 characters, and repeated the evaluation for each interval to comprehensively assess performance across the anticipated range of input lengths. The mean-pooled content embeddings were compressed using PCA to each target dimension. 

The key metric for evaluation was the Recommendation Overlap—the proportion of articles common between the top-$K$ results retrieved from the reduced-dimensional space and the top-$K$ results from the original 768D space.

\paragraph{Results and Analysis for Content Embeddings}
\label{subsub_sec_results_content_dim}

\begin{figure}[h]
  \centering
  \includegraphics[width=1\linewidth, trim=0 0 0 0, clip]{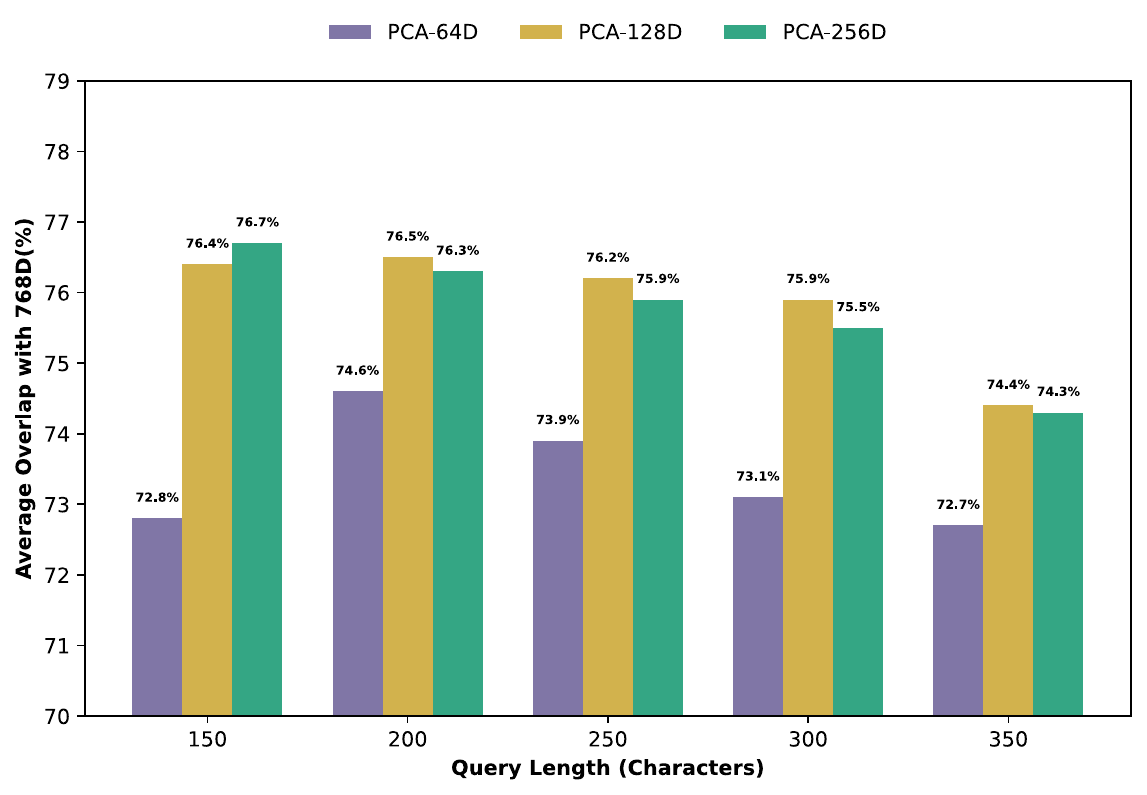}
  \caption{Bar graph analysis of overlap performance comparison across PCA dimensions and query lengths for content embeddings}
  \label{fig:DR_s}
\end{figure}

The performance of PCA-reduced content embeddings across dimensions and query lengths is detailed in Figures ~\ref{fig:DR_s} through grouped bars. The values represent averages computed across the 100 test queries for each length interval. 

A consistent pattern emerged across all query-length scenarios: the 128-dimensional PCA-reduced embeddings achieved the highest mean overlap (75.9\%), outperforming both 64D (73.4\%) and 256D (75.8\%). While 256D embeddings showed marginally higher overlap at the shortest query length (150 characters: 76.7\% vs. 76.4\%), this minimal advantage diminished with longer queries and reversed in the overall mean. The 64D embeddings consistently exhibited the lowest overlap across all intervals.

The results indicate that 128 dimensions provide an effective balance, capturing the essential semantic information of full-length articles with fidelity comparable to higher dimensions while avoiding the unnecessary complexity of 256D. Consequently, 128D was selected as the optimal reduced dimension for content-based retrieval within the long-query pathway.

\subsubsection{Optimal Dimension for Short Queries: Headline Embeddings}
\label{sub_sec_optimal_short}

To determine the optimal reduced dimension for the short-query pathway, we applied a focused evaluation methodology to CLS-pooled headline embeddings. The evaluation considered the inherent characteristics of short queries, which are defined as having fewer than 150 characters and typically exhibit concise, focused semantic intent. Unlike long queries, which span a wide spectrum of lengths from the 150-character threshold to full article content, short queries demonstrate less variability in informational density and structure. Consequently, to determine an optimal dimension that generalizes across the short-query domain, we conducted the evaluation at a single, representative length of 100 characters. This length captures the typical upper bound of a concise search query or headline while remaining well within the short-query threshold, providing a robust benchmark for the compressed semantic space.

A test set of 100 distinct short queries was constructed at the 100-character length. The CLS-pooled headline embeddings were compressed using PCA to 64D, 128D, and 256D, and the performance of each was evaluated against the original 768-dimensional embeddings using the Recommendation Overlap metric, as defined previously.

\paragraph{Results and Analysis for Headline Embeddings}
\label{subsub_sec_results_headline_dim}

\begin{figure}[h]
  \centering
  \includegraphics[width=1\linewidth, trim=50 350 50 60, clip]{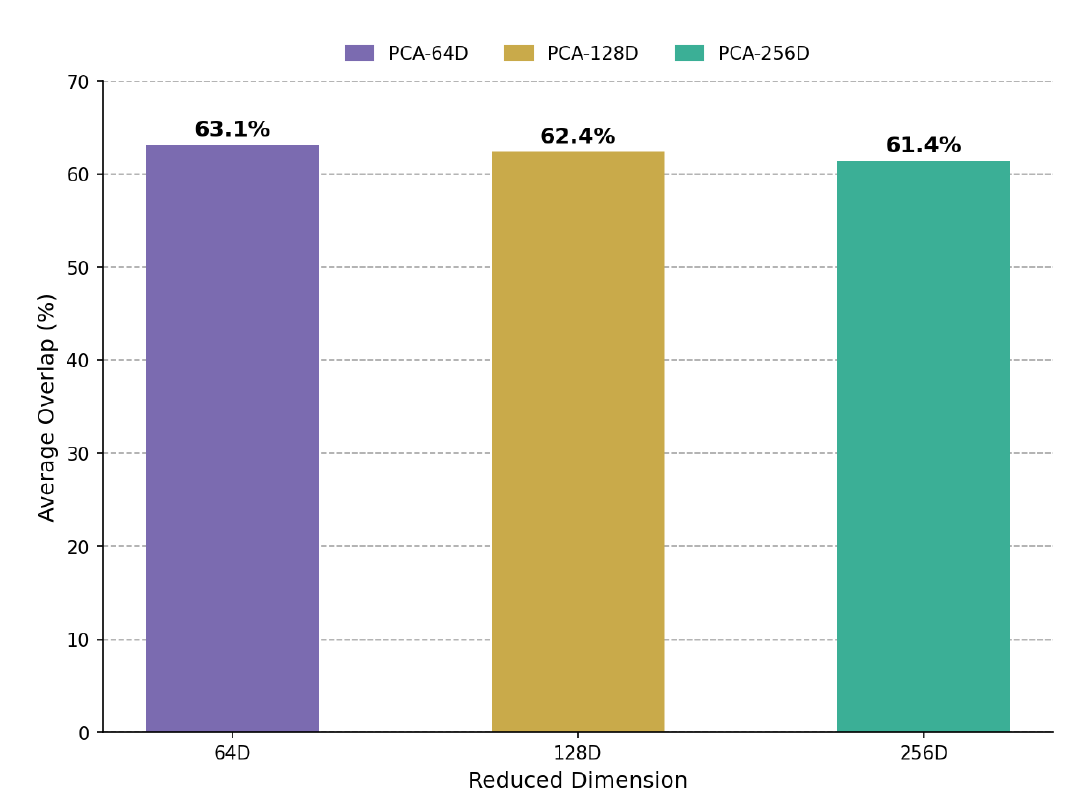}
  \caption{Vertical bar chart showing average overlap percentages for different PCA reduced dimensions (64D, 128D, 256D) using headline embeddings }
  \label{fig:DR_s_h}
\end{figure}

The results, presented in Figure ~\ref{fig:DR_s_h}, reveal a distinct optimum for the headline embedding space. The 64-dimensional PCA-reduced embeddings achieved the highest overlap score of 63.1\%, surpassing both 128D (62.4\%) and 256D (61.4\%). 

This outcome indicates that for the semantically dense and focused space of headlines, a lower-dimensional representation (64D) is not only sufficient but optimal. Effective semantic capture in a smaller vector space is possible due to the concise nature of headlines where core meaning is compressed in fewer tokens. Extending the dimension to 128D or 256D does not offer any benefit and in fact, has a small negative impact on overlap, probably because it adds redundant or noisy dimensions that do not add any significant variance to this particular data distribution.

Therefore 64D was chosen as the best reduced dimension in headline-based retrieval. This decision is consistent with matching the complexity of representations to the complexity of data: simpler, more narrow semantics of headlines can be well represented in a lower-dimensional space, where they can be stored and accessed efficiently, without losing the fidelity needed to do accurate short-query recommendations.

\subsection{Final System Configuration and Performance Evaluation}
\label{sub_sec_final_config}

According to the detailed experimental analysis of the results given in the previous sections, the final retrieval pipeline of the ULTRA framework is as follows:

\begin{itemize}
    \item \textbf{Content Retrieval Pathway (Long Queries, $\ell(q) \geq 150$ }characters\textbf{)}: Embeddings of articles in a mean-pooled representation, downsampled to 128 dimensions with PCA.
    \item \textbf{Headline Retrieval Pathway (Short Queries, $\ell(q) < 150$ }characters\textbf{)}: This uses CLS-pooled headline embeddings reduced to 64 dimensions with PCA.
\end{itemize}

The entire optimized dual-pathway architecture is summarized in Table~\ref{tab:final_config}. The system actively directs queries to the relevant semantic space depending on query length, and each space is optimally relevant using empirically-validated pooling strategies and dimensionality reduction.

\begin{table*}[htbp]
\centering
\caption{Finalized Configuration of the Dual-Pathway Retrieval Pipeline}
\label{tab:final_config}
\begin{tabular}{@{}lcc@{}}
\toprule
\textbf{Component} & \textbf{Content Path (Long Queries)} & \textbf{Headline Path (Short Queries)} \\
\midrule
Source Text & Article Content & Article Headline \\
Pooling Strategy & Mean Pooling & CLS Pooling \\
Dimensionality Reduction & PCA to 128 dimensions & PCA to 64 dimensions \\
Vector Database & ChromaDB (HNSW, cosine space) & ChromaDB (HNSW, cosine space) \\
Key Metadata & Headline, Category & Full Content, Category \\
Optimal Use Case & $\ell(q) \geq 150$ characters & $\ell(q) < 150$ characters \\
\bottomrule
\end{tabular}
\end{table*}

\subsubsection{Final Performance Assessment}
\label{subsub_sec_final_performance}

The ultimate system was tested through a detailed performance evaluation with an extensive test set of 100 queries per pathway to validate the system effectiveness. The true positives (relevant recommendations) were identified using human evaluation, and it was done manually under the strict methodology, as developed in Section~\ref{sub_sec_problemFormulation}. The quantitative measures of Precision@15 show that the system is capable of providing very precise recommendations.

\paragraph{Performance on Short Queries}
Using the headline retrieval pathway which targets brief, keyword-based queries, common in search bars, the system had a mean Precision@15 of 94.35\% on 100 queries. The scores of precision on the 100 test queries were concentrated, with the majority of the test queries having a perfect precision (1.0). This performance is a major success because short queries are a major challenge to semantic matching systems since they are not comprehensive in context. The 94.35\% precision confirms our fundamental hypothesis: matching short queries to semantically compatible headline embeddings is an effective way of bridging the granularity gap that afflicts traditional content-based methods. 

\paragraph{Performance on Long Queries}
In the case of the content retrieval pathway with detailed, context-rich queries or entire article texts, the system registered an outstanding mean Precision@15 of 98.53\%. Although this finding is noteworthy, it is consistent with our experimental findings. The longer queries have more semantic content. Due to this fact, they result in better embedding generation and further similarity matching. The high value of precision confirms that our optimized pipeline, consisting of mean pooling and 128-dimensional PCA reduced embeddings is effective at preserving and matching rich semantic contexts. This performance on a wide range of 100 test queries indicates the strength and stability of this pathway to be implemented in applications like "read-next" article recommendations.

\subsubsection{Discussion of Results}
The difference in the performance between the pathways (94.35\% for short queries and 98.53\% for long queries) indicates the difficulty of each recommendation task. The significant enhancement in the short query case using content embeddings, to 94.35\% (in case of 100 test set queries) using headline embeddings, is the main contribution of our dual-embedding architecture. This change between unacceptable to production-quality performance addresses a serious weakness in semantic recommendation systems. 

The unmatched precision in the long queries, although in the presence of richer input context, however confirms the optimization of all the pipeline components: proper pooling, effective dimensionality reduction, and efficient search of vectors. The outcomes in both directions are an affirmation of the fact that our systematic experimentation and optimization of components has created a robust, high-performance system of recommendation that can support the full range of user query types, both short-term keyword searches to more complex contextual input.

\subsection{Comparative Analysis with Baselines}
\label{sec:comparative_analysis}

It is difficult to create a direct baseline of Urdu content recommendation since the sphere is still young. However, some of the key studies have been the pioneers, and we compare our ULTRA framework with them.

The most applicable published baseline is SEEUNRS (Semantically Enriched Entity-Based Urdu News Recommendation System) \cite{24}. SEEUNRS proposed a two-modular model that included a Semantically Enriched Representation Extractor (SERE) and a Named Entity Extractor (NEE) that used both contextual representations and named entities (persons, places, organizations) to propose relevant articles. On 23,250 news articles curated, the system was tested using the feedback provided by 100 users. Their findings showed that F1 measure was improved by 6.9\% compared to traditional methods including kNN, SVD, and TF-IDF. Although SEEUNRS does not provide a direct precision score, this F1 increase provides a very good semantic-enrichment baseline.

Earlier foundational work consists of the Urdu Wikification framework (2022) \cite{17}, which created an entity-linking pipeline to recognize the named entities and connect them to Wikidata. This paper referenced 16,738 news stories and built a sub-knowledge graph of 8,439 entities. The system combined the TransE-based knowledge graph embeddings with an RNN to attain a recommendation accuracy of 60.8\%. This was the initial evidence of knowledge-enhanced recommendation on Urdu, yet the authors state that the performance is still dependent on the presence of entities in the text.

The other applicable effort is the TF-IDF with BERT embeddings methodology \cite{16}, which used the traditional information retrieval methods on a set of about 1,160 Urdu news articles. With cosine similarity, the system achieved a precision of 0.7143 on news recommendation, with the similarity enhanced with BERT embeddings. Nonetheless, this work is based on a small dataset and fails to consider the variability of query length or semantic granularity.

In a wider context, new developments in the area of Urdu document retrieval have created a good performance standard. A 2026 structure (U-RR²) combining TF-IDF, Word2Vec, and mBERT produced P@10 scores of 0.79 - 0.82 and F1 scores of 0.89 - 0.90 with feature ensemble methods on benchmark collections \cite{53}. Although the results in this case refer to document retrieval as opposed to personalized recommendation, the results demonstrate the feasibility of sophisticated Urdu text processing and provide the validation of  possible performance levels.

In contrast, our ULTRA framework has a Precision@15 of 0.9435 on short, keyword based queries, which is a significant improvement over previously reported values. This is due to our new dual-embedding architecture with length-threshold routing, which has never been thought in any other Urdu recommendation literature. Moreover, a more formidable validation is delivered by our test on a corpus of  around 112,000 articles (orders of magnitude bigger than the 23,250 articles in SEEUNRS) with human-rated relevance scores. Table~\ref{tab:baseline_comparison} summarizes the major differences between ULTRA and other baselines.

\begin{table*}[htbp]
\centering
\caption{Comparison of ULTRA with Urdu Recommendation and Retrieval Systems}
\label{tab:baseline_comparison}
\setlength{\extrarowheight}{8pt} 
\begin{tabular}{@{}p{0.16\linewidth}p{0.34\linewidth}p{0.20\linewidth}p{0.25\linewidth}@{}}
\toprule
\textbf{System} & \textbf{Core Methodology} & \textbf{Reported Performance} & \textbf{Corpus / Dataset} \\
\midrule
TF-IDF + BERT (2022) \cite{16} & TF-IDF + cosine similarity with BERT embeddings & Precision = 0.7143 & $\sim$1,160 articles (Custom Urdu News) \\
\addlinespace 
Urdu Wikification (2022) \cite{17} & Entity linking + RNN with TransE knowledge graph embeddings & 60.8\% accuracy & 16,738 articles (SKRS dataset) \\
\addlinespace
SEEUNRS (2024) \cite{24} & Entity-based semantic extraction (SERE + NEE) with traditional classifiers & 6.9\% F1 improvement over TF-IDF/kNN/SVD & 23,250 articles (100 users) \\
\addlinespace
U-RR² (2026) \cite{53} & Two-stage: (1) TF-IDF + dense embeddings (Word2Vec, FastText, mBERT) + IR models (VSM, BM25, DFR); (2) SVMrank re-ranking with feature ensemble & Weighted Word2Vec: MAP 0.78–0.81, P@10 0.79–0.82; Feature Ensemble: F1 0.89–0.90 & CURE, ROSHNI, UIR\_21 benchmarks \\
\addlinespace
\textbf{ULTRA (This Work)} & \textbf{Dual-Embedding Architecture (Headline CLS-pooling for short queries; Content Mean-pooling for long queries) + PCA} & \textbf{Precision@15 = 0.9435 for short queries} & \textbf{$\sim$118,000 articles} \\
\bottomrule
\end{tabular}
\end{table*}

\section{Conclusions and Future Research Directions}
\label{sec_con_fut_directions}

The current study introduced and validated the ULTRA framework, a new dual-embedding architecture model that seeks to resolve the problem of severe semantic granularity discrepancy in Urdu content ranking. We make our core contribution with a system that dynamically routes queries by length: short, keyword-based queries ($\ell(q) < 150$)  are addressed by CLS-pooled headline embeddings, and long, context-rich queries ($\ell(q) \geq 150$) are handled by mean-pooled content embeddings.

The effectiveness of such approach is conclusively demonstrated by the experimental results. The greatest accomplishment has been to bring the precision of short queries, which is traditionally a difficult task, from an unacceptable and unsatisfactory level to a sturdy 94.35\% (Precision@15) with an evaluation on a larger test set. In the case of long queries, the optimized pipeline scored a remarkable 98.53\% precision, which complimented the selected combination of mean pooling and PCA-based dimensionality reduction to 128 dimensions. The comparative analysis of pooling techniques and dimensionality reduction strategies defined PCA as the best way to retain semantic neighborhoods in both embedding spaces with 64D and 128D as the optimal dimensions of headlines and content paths, respectively.

The work addresses a significant gap in the study of Urdu Natural Language Processing (NLP), in which previous studies have mainly dealt with text classification and summarization as foundational tasks. We offer a validated, high performance content discovery architecture of Urdu by creating and rigorously testing an advanced semantic recommendation system.

The effectiveness of this framework provides multiple avenues of future work, not only to solve its shortcomings but also to enlarge its possibilities. The nearest future move is the implementation of this system as a commercial application. The next logical step is to incorporate the ULTRA framework into a specific Urdu news portal or aggregator. This app may have a simple search interface to support brief searches and offer the long-query pathway-driven "Read Next" or "Related Articles" panels at the bottom of articles.

Moreover, the recommendation engine could be packaged as a cloud-based API service. This would allow existing Urdu digital publishers, blogging platforms, or digital libraries to integrate sophisticated semantic recommendation capabilities into their websites without developing the core NLP infrastructure in-house.
The system can also be evolved into a hybrid recommender. The current semantic component could be combined with a collaborative filtering module that analyzes anonymized user interaction data. The final recommendation score could be a weighted combination of semantic similarity and personalized user relevance. 
Another interesting direction could be figuring out the same dual-embedding logic be applied to recommend related English articles for an Urdu query, or vice-versa, by using a multilingual embedding model. Furthermore, with the rise of multimedia news, could the system be extended to recommend relevant images or short video clips based on the semantic content of an article. 

\section*{Acknowledgment}
The authors acknowledge the use of the DeepSeek AI assistant as a supportive tool during the composition of this manuscript. It was utilized for sentence structuring, translation of ideas, and ensuring grammatical and formal English language generation, thereby aiding in the clear and effective communication of the research findings.

\begin{IEEEbiography}[{{\includegraphics[width=1.1in,height=1.35in,keepaspectratio]{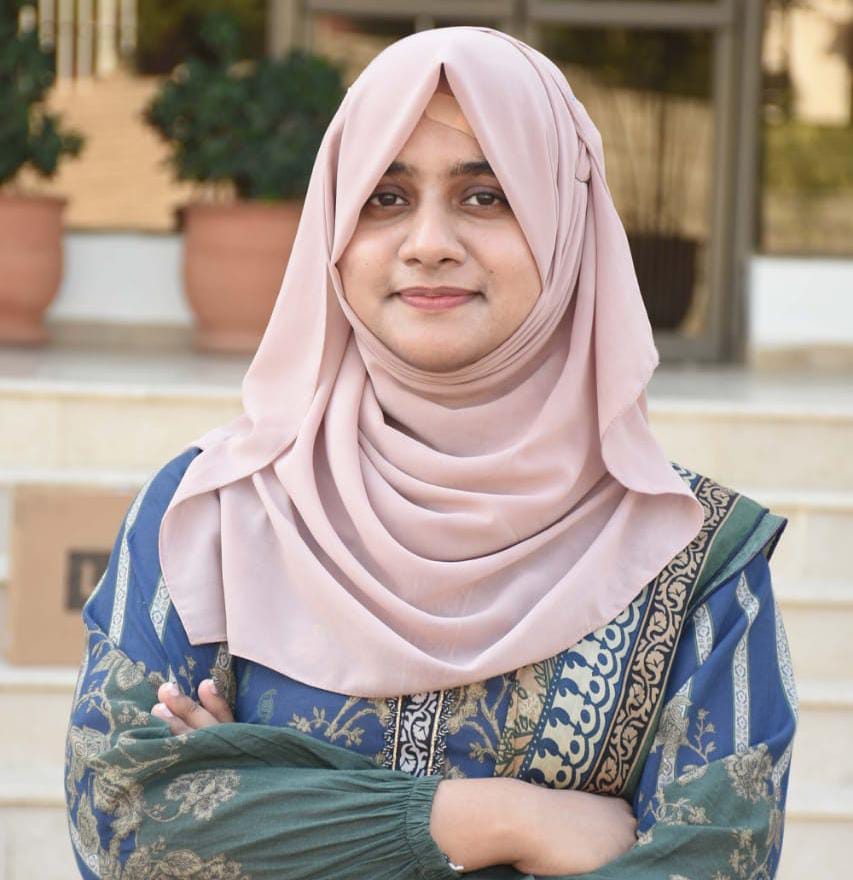}}
}]{Alishbah Bashir}
is currently pursuing her B.S. degree in Computer and Information Sciences at the Pakistan Institute of Engineering and Applied Sciences (PIEAS), Islamabad, Pakistan. Her research interests are centered on Artificial Intelligence (AI) and its applications, with a particular focus on Retrieval-Augmented Generation (RAG), Intelligent Information Retrieval, Recommender Systems, Agentic AI, Natural Language Processing (NLP), and the fine-tuning of LLMs.
\end{IEEEbiography}

\begin{IEEEbiography}[{\includegraphics[width=1in,height=1.25in,clip,keepaspectratio]{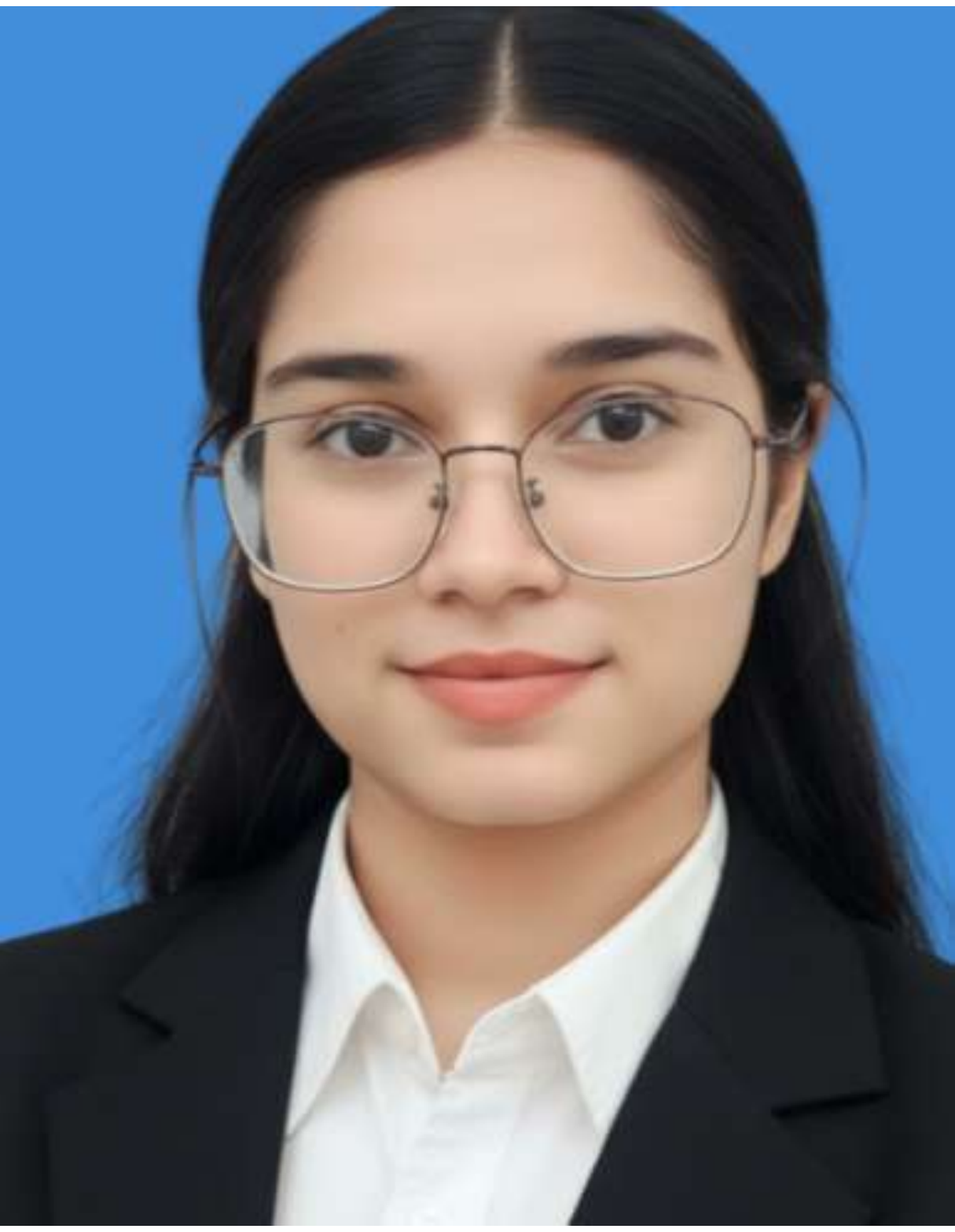}}]{ Fatima Qaiser} is an undergraduate student in the Department of Computer and Information Sciences at PIEAS, Islamabad. Her research focuses on large language models (LLMs), agentic AI systems, and natural language processing (NLP), with an emphasis on reliability, interpretability, and fairness in language-centric models. She works on the analysis and design of LLM-based agents, studying reasoning consistency, transparency, and robustness in autonomous and semi-autonomous decision-making settings. Her interests include explainability methods for LLMs, agentic reasoning pipelines, and NLP-driven architectures for trustworthy AI applications.
\end{IEEEbiography}

\begin{IEEEbiography}[{\includegraphics[width=1in,height=1.25in,clip,keepaspectratio]{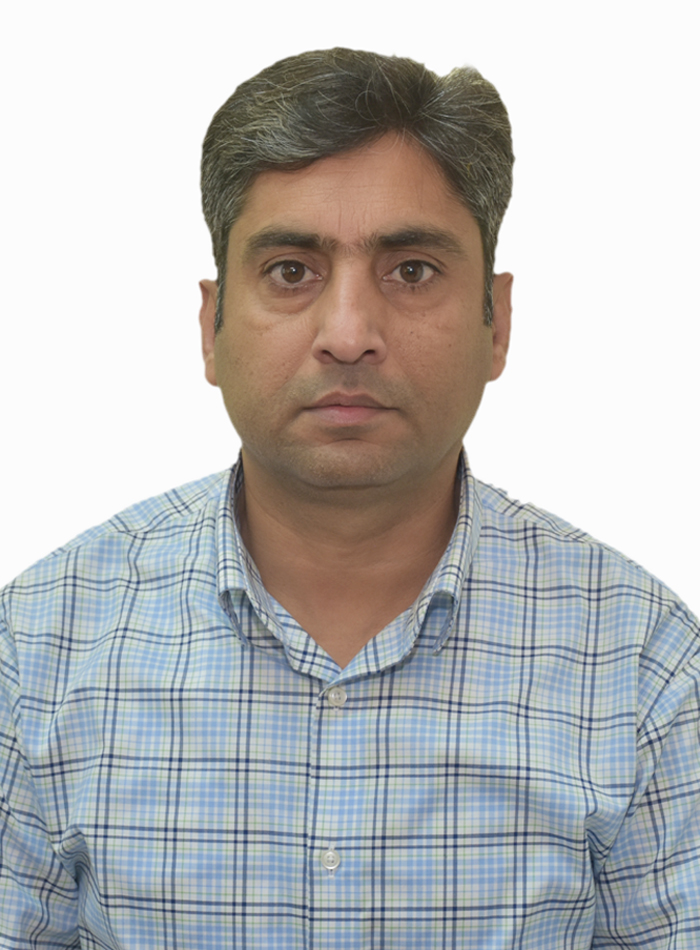}}]{IIjaz Hussain} received his Ph.D. in Computer Science from COMSATS University Islamabad, where he built a solid foundation in computational theories. He currently serves as a Associate Professor in the Department of Computer and Information Sciences at the Pakistan Institute of Engineering and Applied Sciences (PIEAS), Islamabad. He has authored over twenty research articles in reputable journals, and presented at international conferences, contributing significantly to the field. His research interests include natural language processing (NLP), information retrieval, and the application of NLP in cybersecurity.  
\end{IEEEbiography}

\EOD

\end{document}